\documentclass[twocolumn,showpacs,aps,prd,groupedaddresss,letterpaper]{revtex4}

\usepackage{graphicx}
\usepackage{epsfig}
\usepackage{amsmath}
\usepackage{amssymb}
\usepackage{hhline}
\usepackage{multirow}
\usepackage{fancyheadings}
\usepackage{ifthen}
\usepackage{lscape}
\usepackage{rotating}
\usepackage{here}
\usepackage{wrapfig}
\usepackage{color}
\usepackage{dcolumn}
\RequirePackage{xspace}

\newcommand{\tev}{\ensuremath{\mathrm{\,Te\kern -0.1em V}}\xspace}
\newcommand{\gev}{\ensuremath{\mathrm{\,Ge\kern -0.1em V}}\xspace}
\newcommand{\gevv}{\ensuremath{{\mathrm{\,Ge\kern -0.1em V}^2}}\xspace}
\newcommand{\gevvv}{\ensuremath{{\mathrm{\,Ge\kern -0.1em V}^4}}\xspace}
\newcommand{\gevn}{\ensuremath{{\mathrm{\,Ge\kern -0.1em V}^n}}\xspace}
\newcommand{\mev}{\ensuremath{\mathrm{\,Me\kern -0.1em V}}\xspace}
\newcommand{\kev}{\ensuremath{\mathrm{\,ke\kern -0.1em V}}\xspace}
\newcommand{\ev}{\ensuremath{\mathrm{\,e\kern -0.1em V}}\xspace}
\newcommand{\gevc}{\ensuremath{{\mathrm{\,Ge\kern -0.1em V\!/}c}}\xspace}
\newcommand{\mevc}{\ensuremath{{\mathrm{\,Me\kern -0.1em V\!/}c}}\xspace}
\newcommand{\gevcc}{\ensuremath{{\mathrm{\,Ge\kern -0.1em V\!/}c^2}}\xspace}
\newcommand{\mevcc}{\ensuremath{{\mathrm{\,Me\kern -0.1em V\!/}c^2}}\xspace}

\def\Vub  {\ensuremath{|V_{ub}|}\xspace}
\def\Vcb  {\ensuremath{|V_{cb}|}\xspace}

\def\ra                 {\ensuremath{\rightarrow}\xspace}
\def\to                 {\ensuremath{\rightarrow}\xspace}

\usepackage{relsize}
\def\babar{\mbox{\slshape B\kern-0.1em{\smaller A}\kern-0.1em
    B\kern-0.1em{\smaller A\kern-0.2em R}}}

\newcommand{\verysmallrule}{\rule[-2.0mm]{0.0cm}{0.8cm}}
\newcommand{\vvsmallrule}{\rule[-1.0mm]{0.0cm}{0.6cm}}

\def\ecut {\ensuremath{E_{\rm cut}}}
\def\mmx {\ensuremath{\langle M_X \rangle}}
\def\mmxd {\ensuremath{\langle M_X^3 \rangle}}

\def\mmxc {\ensuremath{\langle M_X^3 \rangle}}

\def\mmxn {\ensuremath{\langle M_X^n \rangle}}
\def\meln {\ensuremath{\langle E_{\ell}^n \rangle}}

\def\megn {\ensuremath{\langle E_{\gamma}^n \rangle}}
\def\meg {\ensuremath{\langle E_{\gamma} \rangle}}

\def\vegs {\ensuremath{\langle (E_{\gamma} -\meg)^2 \rangle}}

\def\vcb {\ensuremath{|V_{cb}|}}

\def\gsl {\ensuremath{\Gamma_{\rm SL}}}

\def\msb {\ensuremath{\overline{\mathrm{MS}}}}
\def\btoclv{\ensuremath{B \ra X_c \ell \bar{\nu}}}
\def\btosg{\ensuremath{B \ra X_s \gamma}}

\newcommand{\GeV}{\,\mbox{GeV}}

\newcommand{\msp}[1]{\mbox{\hspace*{#1mm}~}}

\long\def\inst#1{\par\nobreak\kern 4pt\nobreak
    {\it #1}\par\vskip 10pt plus 3pt minus 3pt}

\begin{document}


\title[]{{\large \bf \boldmath Fit to Moments of Inclusive \btoclv\ and \btosg\ Decay Distributions\\ 
using Heavy Quark Expansions in the Kinetic Scheme}}

\author{O.~L.~Buchm\"{u}ller}
\affiliation{CERN, CH-1211 Geneva 23, Switzerland}
\author{H.~U.~Fl\"{a}cher}
\affiliation{Royal Holloway, University of London, Egham, Surrey TW20 0EX, United Kingdom}

\begin{abstract}

We present a fit to measured moments of inclusive 
distributions in \btoclv\ and \btosg\ decays to extract values for the CKM matrix element \Vcb , 
the b- and c- quark masses, and higher order parameters that appear in the Heavy Quark Expansion. 
The fit is carried out using theoretical calculations in the kinetic scheme and includes moment 
measurements of the \babar , Belle, CDF, CLEO and DELPHI collaborations for which correlation 
matrices have been published.
We find $\Vcb = ( 41.96 \pm 0.23_{\rm exp} \pm 0.35_{\rm HQE} \pm 0.59_{\gsl} ) \times 10^{-3}$ and   
$m_b = 4.590 \pm 0.025_{\rm exp} \pm 0.030_{\rm HQE} \gev$ where the errors are experimental and 
theoretical respectively.
We  also derive values for the heavy quark distribution function parameters $m_b$ and $\mu_{\pi}^2$ 
in different theoretical schemes that can be used as input for the determination of \Vub. 

\end{abstract}


\date{June 4, 2006}

\maketitle

\section{Introduction}

In the past few years tremendous progress has been made in the description of semileptonic and radiative $B$ decays using the framework of Heavy Quark Expansions (HQEs). 
Calculations for the semileptonic decay width as well as for moments of inclusive observables with restrictions on the phase space
are now available in different schemes through order 
$1/m_b^3$ and $\alpha_s^2 \beta_0$
~\cite{Falk:1997jq,KolyaPaolo,Kolya,Benson:2004sg,Bauer:2002sh,Uraltsev:2004in}. 
At the same time many new experimental measurements of moments of the hadronic mass and lepton energy distribution in \btoclv\ as 
well as the photon energy spectrum in \btosg\ decays have been carried out by several experiments. Generally, the results agree 
very well between experiments.
In addition, the theoretical calculations describe the measured data well establishing this framework for treating semileptonic 
and radiative $B$ decays~\cite{BABAR,Bauer:2004ve,DELPHI}.\\

In this document we will present the results of a combined fit to 
measured moments for which correlation matrices are published.
These include moments of the hadronic mass distribution \mmxn\ and moments of the lepton energy 
spectrum \meln\ in inclusive \btoclv\ decays as well as moments of the photon energy spectrum \megn\ in 
inclusive \btosg\ decays for different minimum lepton and photon energies $E_{\rm cut}$.
The HQEs for the moments depend on the b- and c- quark masses and several non-perturbative 
parameters which therefore can be determined from a fit of the theoretical expressions to the 
experimental moment measurements.
Among the measurements we have excluded those for which there are ongoing discussions 
within the theoretical community regarding the associated theoretical uncertainties. 
These are in particular the non-integer moments \mmx\ and \mmxc\ of the hadronic mass distribution 
in \btoclv\ decays. 
In addition, 
moments of the photon energy spectrum in \btosg\ decays above $E_{\rm cut} = 2.0 \gev$ have been excluded
since there the standard local Operator Product Expansion (OPE) is no longer believed to be under theoretical control.
Furthermore, individual moment measurements have been discarded if the covariance matrix cannot be inverted
due to the large correlations between the measurements.

\section{Heavy Quark Expansions in the Kinetic Scheme}
\label{sec:theory}

 In this analysis we make use of HQEs 
that express the semileptonic decay width 
$\Gamma_{\rm SL}$ as well as moments of the lepton energy and hadron mass distribution in 
\btoclv\ decays and those of the photon energy spectrum in \btosg\ decays in terms of the running kinetic quark masses $m_b(\mu)$ and $m_c(\mu)$. 
Non-perturbative effects are introduced in this formalism 
via heavy quark operators. 
The leading power corrections
arise at $O(1/m_b^2)$ and are controlled by the two expectation values $\mu_{\pi}^2(\mu)$ 
and $\mu_{G}^2(\mu)$ of the kinetic and chromomagnetic dimension-five operators. 
At $O(1/m_b^3)$ two additional expectation values 
$\rho_{D}^3(\mu)$ and $\rho_{LS}^3(\mu)$ of the Darwin and spin-orbital (LS) 
dimension-six operators complete
the set of non-perturbative corrections. Together with the two running quark masses the HQE
in the kinetic scheme includes six free parameter through $O(1/m_b^3)$:
\begin{itemize}
\item {Leading Order Parameters}
\begin{itemize}
\item $m_b(\mu)$ $\rightarrow$ b-quark mass
\item $m_c(\mu)$ $\rightarrow$ c-quark mass
\end{itemize}
\item {Leading Non-Perturbative Corrections - $O(1/m_b^2)$}
\begin{itemize}
\item $\mu_{\pi}^2(\mu)$ $\rightarrow$ `kinetic expectation value'
\item $\mu_{G}^2(\mu)$ $\rightarrow$ `chromomagnetic expectation value'
\end{itemize}
\item {Higher Order Non-Perturbative Corrections - $O(1/m_b^3)$}
\begin{itemize}
\item $\rho_{D}^3(\mu)$  $\rightarrow$ `Darwin term'
\item $\rho_{LS}^3(\mu)$ $\rightarrow$ `spin-orbital term'
\end{itemize}
\end{itemize}   
At any given value of the 
Wilsonian factorization scale $\mu$ all of the above mentioned HQE
parameters represent well defined physical quantities which have to be determined from measurements. 
The scale $\mu$ separates `short-distance' effects from `long-distance' effects and, therefore,
$m_Q(\mu)$ can be understood as a short-distance mass of perturbation theory that excludes soft gluon 
interactions~\cite{Kolya}. It is important to note that Ref.~\cite{Kolya} contains a translation to full order $\alpha_s^2$ and third-order
BLM corrections of the running short-distance mass $m_Q(\mu)$ into the well known running \msb\ mass $\overline{m}_Q(\overline{m}_Q)$. 
Hence the kinetic masses $m_b(\mu)$ and $m_c(\mu)$ can be compared with other established
mass definitions in QCD such as the \msb\ mass. 
This fact is important for
the comparison with other QCD calculations beyond semileptonic or rare B-decays.
In order to minimize the influence of radiative corrections and to insure that the 
kinetic c-quark mass $m_c(\mu)$ has a well defined physical meaning
the separation scale is set to be $\mu=1~\gev$. A detailed
discussion of the justification for this choice can be found in Ref.~\cite{scale}. In the 
following, all heavy quark parameter values are presented for $\mu=1~\gev$ (e.g. $m_b(1~\gev) \rightarrow m_b$). 
The analytical expression for the semileptonic width of
$B\rightarrow X_c \ell\bar{\nu}$ decays through $O(1/m_b^3)$~\cite{Kolya} is given by
\begin{eqnarray}
\label{equ:opegsl}
\nonumber
\lefteqn{\Gamma_{\rm SL}(\btoclv) \msp{-1} =} & & \msp{25}  \frac{G_F^2 m_b^5}{192\pi^3} |V_{cb}|^2 (1+A_{ew}) A_{pert}(r,\mu)  \nonumber \\
& & \times \Bigg [ z_0(r) \Bigg ( 1 - \frac{\mu_{\pi}^2-\mu_G^2+\frac{\rho_D^3+\rho_{LS}^3}{m_b}}{2m_b^2} \Bigg ) \\
& & \msp{3} - 2(1-r)^4\frac{\mu_G^2+\frac{\rho_D^3+\rho_{LS}^3}{m_b}}{m_b^2}  +d(r)\frac{\rho_D^3}{m_b^3}+O(1/m_b^4)\Bigg],  \nonumber
\end{eqnarray}
where $r=m_c^2/m_b^2$ explicitly contains the c-quark mass $m_c$. The tree level phase space factor $z_0$ is defined
through:
\[ z_0(r)=1-8r+8r^3-r^4-12r^2\mathrm{ln}(r)\quad ,\]
and the expression $d(r)$ is given by:
\[ d(r)=8\mathrm{ln}(r)+\frac{34}{3}-\frac{32}{3}r-8r^2+\frac{32}{3}r^3-\frac{10}{3}r^4 \quad . \]
The electroweak corrections for Eq.~\ref{equ:opegsl} are estimated to be approximately
\[ 1+A_{ew} \approx \bigg ( 1 + \frac{\alpha}{\pi} \mathrm{ln}\frac{M_Z}{m_b} \bigg)^2\approx 1.014 \quad , \]  
and the perturbative contributions, which have been calculated to all orders in BLM corrections and to
second order in non-BLM corrections, are for a reasonable set of
HQE parameters estimated to be $A_{pert}\approx 0.908$. 
It should be noted that Eq.~\ref{equ:opegsl} is not a HQE in powers of $1/m_b$.
Since the most relevant scale for the $b\rightarrow c$ transition is the energy release $m_b-m_c$ of the decay rather than the b-quark mass $m_b$, 
this expansion is carried out in powers of $1/(m_b-m_c)$.
On the other hand, due to the low mass of the $s-$ and $u-$quark, HQE calculations for the $b\rightarrow s$ or $b\rightarrow u$ 
transition can be considered as expansions in powers of $1/m_b$.
The full $\alpha_s^2$ corrections to the
photon spectrum have been computed recently~\cite{Melnikov:2005bx}. 
As follows from this result, the effects of omitting the non-BLM corrections are small 
and fully covered by the assumed theoretical uncertainties we
quote for the first and second photon energy moments.
\\ 

For the practical use Eq.~\ref{equ:opegsl}
can be transformed into:
{\small
\begin{eqnarray}
\label{equ:opevcb}
\nonumber
\lefteqn{\frac{|V_{cb}|}{|V_{cb}^0|} \msp{0} =} & & \msp{11} \Big ( \frac{BR_{c\ell\bar{\nu}}}{0.105 - 0.0018} \Big )^{\frac{1}{2}}
\Big (\frac{1.55\mathrm{ps}}{\tau_{B}} \Big  )^{\frac{1}{2}} 
(1 \pm \delta _{\rm th}) \\ 
& & \times \,[1+0.30\,(\alpha_s(m_b)\!-\!0.22)]\,     \\
& & \times \left[1 -0.66 \left ( m_b \!-\!4.6\GeV \right )
+0.39 \left ( m_c \!-\!1.15\GeV \right )\qquad \right. \nonumber \\
& &  \msp{2}+0.013 \left ( \mu_{\pi}^2 \!-\!0.4\GeV^2 \right )
+0.09 \left ( \tilde{\rho_D}^3 \!-\!0.1\GeV^3 \right ) \nonumber \\
& & \msp{2}\left. +\,0.05 \left ( \mu_G^2\!-\!0.35\GeV^2 \right ) 
-0.01 \left ( \rho_{LS}^3 \!+\!0.15\GeV^3 \right ) \right ]\;, \nonumber
\end{eqnarray}
}
where $|V_{cb}^0| = 0.0417$ and
$\tau_B$ represents the average lifetime of neutral and charged B mesons.
We use $\tau_B = 1.585 \pm 0.007$ ps based on Ref.~\cite{PDG2005}, assuming equal production of charged and neutral $B$ mesons. 
The theoretical uncertainty due to the the limited accuracy of this HQE is denoted by $\delta_{\rm th}$. 
In Ref.~\cite{Kolya} its value is quoted with $\delta_{\rm th}=0.015$, with contributions from four different sources. 
Since then a more elaborate study on the influence of `intrinsic charm' has been carried out reducing the associated uncertainty by roughly a factor two~\cite{Bigi:2005bh}, leaving an overall uncertainty of $\delta_{\rm th}=0.014$.
More details about this theoretical uncertainty can be found in Section~\ref{theory_errors}.\\
It should be noted that we do not extract the Darwin term $\rho_{D}^3(1~\gev)$ directly from our fit to the HQEs but rather the
`pole-type' Darwin expectation value $\tilde{\rho_{D}}^3$. The two parameters are closely related in our framework via
\[ \tilde{\rho_{D}}^3 = \rho_{D}^3(1 \gev) - 0.1 \gev^3 \quad .\]

\begin{table*}[hbt]
\caption[]{\label{tab:momsum} Summary of moment measurements used in the combined fit. $n$ indicates the order of the (central) moment measurement of observable $\langle M_X^n \rangle_{E_{\rm cut}}$, $\langle E_{\ell}^n \rangle_{E_{\rm cut}}$ and $\langle E_{\gamma}^n \rangle_{E_{\rm cut}}$. $E_{\rm cut}$ indicates measurements with the corresponding minimum lepton momenta and photon energies in \gev. }
\begin{center}
\begin{tabular}{c |cl |cl |cl}\hline\hline
        \verysmallrule Experiment & \multicolumn{2}{c|}{Hadron Moments $\langle M_X^n \rangle_{E_{\rm cut}}$} & \multicolumn{2}{c|}{Lepton Moments $\langle E_{\ell}^n \rangle_{E_{\rm cut}}$} & \multicolumn{2}{c}{Photon Moments $\langle E_{\gamma}^n \rangle_{E_{\rm cut}}$}\\\hline\hline	
\babar & $n$=2 &$E_{\rm cut}$=0.9,1.0,1.1,1.2,1.3,1.4,1.5     & $n$=0& $E_{\rm cut}$=0.6,1.2,1.5          & $n$=1& $E_{\rm cut}$=1.9,2.0 \footnotemark[1]  \\
\cite{BABARHAD,BABARLEP}       & $n$=4 & $E_{\rm cut}$=0.9,1.0,1.1,1.2,1.3,1.4,1.5     & $n$=1& $E_{\rm cut}$=0.6,0.8,1.0,1.2,1.5  & $n$=2& $E_{\rm cut}$=1.9 \footnotemark[1] \\
\cite{BABARSEMI,BABARINCL} &            &                                    & $n$=2& $E_{\rm cut}$=0.6,1.0,1.5  &&  \\
       &             &                                   & $n$=3& $E_{\rm cut}$=0.8,1.2  &  &\\\hline
Belle  &       &     &    && $n$=1& $E_{\rm cut}$=1.8,1.9           \\   
\cite{BELLEBSG,BELLEBSGNEW}     &       &      &  & & $n$=2&   $E_{\rm cut}$=1.8,2.0      \\\hline
CDF    & $n$=2 & $E_{\rm cut}$=0.7         &  & &                 &\\
\cite{CDF}       & $n$=4 & $E_{\rm cut}$=0.7      &   &&                &\\\hline
CLEO   & $n$=2&$E_{\rm cut}$=1.0,1.5         & &         & $n$=1 & $E_{\rm cut}$=2.0           \\   
\cite{CLEOHAD,CLEOBSG}       & $n$=4&$E_{\rm cut}$=1.0,1.5  & &   &     \\\hline
DELPHI & $n$=2&$E_{\rm cut}$=0.0      & $n$=1&$E_{\rm cut}$=0.0     &  &\\   
\cite{DELPHI}       & $n$=4&$E_{\rm cut}$=0.0      & $n$=2&$E_{\rm cut}$=0.0    & &\\
       & $n$=6&$E_{\rm cut}$=0.0      & $n$=3&$E_{\rm cut}$=0.0     & &\\\hline
HFAG \cite{HFAG2004}  &      &               & $n$=0&$E_{\rm cut}$=0.6    & & \\ 
	\hline\hline
	\end{tabular}
\footnotetext[1]{A total of six photon moments from Refs.~\cite{BABARSEMI} and~\cite{BABARINCL} are used.}
\end{center}
\end{table*}

Relations similar to Eq.~\ref{equ:opevcb} have been calculated for inclusive observables in semileptonic and radiative $B$ decays. These are in particular
hadron mass \mmxn\ and lepton energy moments \meln\ to order $n$ as well as moments of the photon energy spectrum \megn:
\begin{eqnarray}
\nonumber
\label{equ:opemom}
\mmxn & \rightarrow & \mmxn \big (m_b,m_c,\mu_{\pi}^2,\mu_{G}^2,\rho_{D}^3,\rho_{LS}^3,\alpha_s\big ) \nonumber\\
\meln & \rightarrow & \meln \big (m_b,m_c,\mu_{\pi}^2,\mu_{G}^2,\rho_{D}^3,\rho_{LS}^3,\alpha_s\big )\nonumber \\
\megn & \rightarrow & \megn \big (m_b,\mu_{\pi}^2,\mu_{G}^2,\rho_{D}^3,\rho_{LS}^3,\alpha_s\big ) \quad . \nonumber
\end{eqnarray} 
Since every moment calculation has a different dependence on the heavy quark parameters
a simultaneous fit allows for the extraction of all these parameters.
For this it is important to use as many moment measurements as possible in order to overconstrain the extraction of the heavy quark parameters and 
to establish the validity of the expansions.
A much more detailed description of the theoretical framework used for this analysis can be found in Refs.~\cite{KolyaPaolo,Kolya,Benson:2004sg}.

\section{Fit to Moment Measurements}
\label{fitresult}

In the following sections we list the currently available experimental moment measurements and indicate which are used in the combined fit presented here.
Furthermore, we outline the fit procedure and summarize the results obtained from different fit scenarios and various cross checks. 

\subsection{Experimental Input}
\label{exp_input}
All results are based on the following set of moment measurements which are also summarised in Table~\ref{tab:momsum}. 
Additional measurements for which correlation matrices are not available and thus cannot be used in the presented fit are listed in parentheses.
\begin{itemize}
\item \babar\\
       Hadron mass~\cite{BABARHAD} and lepton energy moments~\cite{BABARLEP} from \btoclv\ decays measured as a function of the minimum lepton energy $E_{\rm cut}$. The lepton moments used here differ slightly from those in the \babar\ publication~\cite{BABARLEP}. 
They have been updated by taking into account the recent improved measurements of the $D_s$ and $B$ branching fractions ({\it upper-vertex charm}) that impact the background subtraction.
      Moments of the photon energy spectrum in \btosg\ decays as a function of the minimum photon energy $E_{\rm cut}$ from two independent analyses~\cite{BABARSEMI,BABARINCL}.
\item Belle\\
      First and second moment of the photon energy spectrum as a function of the minimum photon energy $E_{\rm cut}$~\cite{BELLEBSG,BELLEBSGNEW}.\\
(Measurements of hadron mass and lepton energy moments as functions of the lower lepton energy exist~\cite{BELLEHAD,BELLELEP} but are excluded from the current fit as correlation matrices are only available for the statistical errors.)  
\item CDF\\
      Hadron moment measurements with a minimum lepton energy of $E_{\rm cut}=0.7$ \gev\ \cite{CDF}.
\item CLEO\\
      Hadron moment measurements as a function of the minimum lepton energy \cite{CLEOHAD}.\\
      First (and second) moment of the photon energy spectrum at $E_{\rm cut}=2.0$ \gev~\cite{CLEOBSG}.\\
      (The measurement of lepton energy moments as a function of  $E_{\rm cut}$~\cite{CLEOLEP} is not 
       given with the full covariance matrix and thus has not been included in the fit~\cite{stepaniak}.)
\item DELPHI\\
      Lepton energy and hadron mass moment measurements with no restriction on the lepton energy~\cite{DELPHI}.	
\end{itemize}

\subsection{Fit Procedure}
A $\chi^2$ minimization technique is used to determine the HQE predictions in the kinetic scheme from a fit to 
the data: 
\begin{equation}
\label{equ:chi}
\chi^2 = (\vec{M}_{\rm exp} - \vec{M}_{\rm HQE})^T  C_{\rm tot}^{-1} (\vec{M}_{\rm exp} - \vec{M}_{\rm HQE})
\end{equation}
where $\vec{M}_{exp}$ represents all moment measurements included in the fit and $\vec{M}_{\rm HQE}$ stands 
for their HQE prediction defined by the heavy quark parameters. $C_{\rm tot}=C_{\rm exp}+C_{\rm theo}$ is 
the sum of the experimental and theoretical covariance matrices. The construction of the theoretical 
covariance matrix $C_{\rm theo}$ is discussed in detail in Section~\ref{theory_errors}.\\

In order to extract the semileptonic branching fraction for \btoclv\ events, 
$BR_{c\ell\bar{\nu}}$, from the fit, the measurements of the truncated branching fractions 
$\langle E^0_L \rangle$ at different cutoffs in the lepton energy $E_{\rm cut}$ are also used 
as experimental input. 
HQE predictions of the relative decay fraction for a given cutoff $E_{\rm cut}$,
\begin{equation}
\label{eqy:R}
R(E_{\rm cut}) = \frac{\int_{E_{\rm cut}} \frac{d\Gamma}{dE_L} dE_L}{\int_0 \frac{d\Gamma}{dE_L} dE_L}
\end{equation}
can be used to extrapolate the measurement of the truncated branching fractions to zero cutoff:
\begin{equation}
\label{eqy:BR}
BR_{c\ell\bar{\nu}} = \frac{\langle E^0_L\rangle_{E_{\rm cut}}}{R(E_{\rm cut})} \quad . 
\end{equation}
In addition to the \babar\ measurement we also include the HFAG average at $E_{\rm cut} = 0.6 \gev$ of 
$(10.29 \pm 0.18) \%$~\cite{HFAG2004}.  
In order to utilize more than only one of these truncated branching fractions, one has to include the 
total branching fraction $BR_{c\ell\bar{\nu}}$ as a free parameter in the fit. 
Together with the input of the 
averaged $B$ meson
lifetime $\tau_{B}= 1.585 \pm 0.007$ ps  
this can be used to calculate the semileptonic width $\Gamma_{\rm SL}^{c\ell\bar{\nu}}$ as part of the $\chi^2$ minimization. 
The HQE for $\Gamma_{\rm SL}^{c\ell\bar{\nu}}$ is directly related to $|V_{cb}|^2$ (see Eq.~\ref{equ:opegsl})
and introducing $|V_{cb}|$ as a free parameter therefore has the advantage of determining the error on this quantity directly from the global fit. 
With this approach also potential non-Gaussian errors ({\it e.g.} asymmetric errors) of the fit parameters are properly propagated into the error on $|V_{cb}|$.\\

The fit 
to the moment measurements is carried out using the HQE calculations in the kinetic scheme presented 
in Refs.~\cite{KolyaPaolo,Benson:2004sg}, including $E_{\rm cut}$ dependent perturbative corrections 
to the hadron moments~\cite{Uraltsev:2004in,Trott:2004xc,Aquila:2005hq}.
However, rather than using linearized tables to determine $\vec{M_{HQE}}$ 
(as was done previously in Ref.~\cite{BABAR})
, we obtain the prediction for every single moment from an analytical calculation~\cite{Fortran}.
This not only allows us to study the scale dependence of the kinetic scheme in detail 
but also provides more accurate predictions for the individual moments. 
Since some criticism has recently been raised concerning the quality of the 
theoretical expansion for the non-integer hadron moments \mmx\ and \mmxd, we exclude these moments from
the fit until the issue of the theoretical uncertainty of these moment predictions is resolved. 
However, for comparison, we will always show their prediction based on the fit results and compare them with the corresponding measurements.

As $\mu_{G}^2$ and $\rho_{LS}^3$ are estimated from the $B^* - B$ mass 
splitting and heavy-quark sum rules, respectively, we impose Gaussian error 
constraints of $\mu_{G}^2=0.35\pm0.07 \gev^2$ and $\rho_{LS}^3=-0.15\pm 0.10 \gev^3$ on these parameters as advocated in Ref.~\cite{KolyaPaolo}.

\subsection{Theoretical Error Estimates}
\label{theory_errors}
Since this analysis targets a measurement of the CKM matrix element $|V_{cb}|$ with a relative error below the 2\% level 
it is of vital importance to take theoretical uncertainties 
into account, as currently estimated.
The HQEs for the moments have
theoretical uncertainties due to 
certain limitations in the accuracy of the 
calculations and certain approximations.
As Eq.~\ref{equ:opevcb} illustrates, the heavy quark parameters extracted from the simultaneous fit
to the moments are used for a residual correction to $|V_{cb}^0|$.
It is therefore important to include their theoretical uncertainties in the total covariance matrix 
of the $\chi^2$ fit to achieve realistic error estimates for them and for \Vcb.

Since the HQE for $\Gamma_{\rm SL}$ is the best known expansion in this framework also its error estimates are the most advanced.
In Eq.~\ref{equ:opevcb} the HQE for $|V_{cb}|$ is explicitly quoted with a theoretical uncertainty $\delta_{th}$
to reflect the limited accuracy of this theoretical expression. Adding the individual uncertainties 
quoted in Ref.~\cite{Kolya} in quadrature and accounting for the more advanced estimates of the potential importance of `intrinsic charm' 
of Ref.~\cite{Bigi:2005bh},
yields $\delta_{th}=1.4\%$. 
We quote this error separately for $|V_{cb}|$ and label it with a subscript `$\gsl$'. \\

Following the recipe quoted in Ref.~\cite{KolyaPaolo} we estimate the uncertainties in the individual 
HQEs for the moments $M$ by conservatively varying the values of $\mu_{\pi}^2$ and 
$\mu_G^2$ by $\pm 20\%$ and those of the $O(1/m_b^3)$ operators $\rho_{LS}^3$ and 
$\rho_D^3$ by $\pm 30 \%$. 
These variations are carried out around the expected theoretical values of  $\mu_{\pi}^2 = 0.4 \gev^2$, 
$\mu_G^2 = 0.35 \gev^2$,  $\rho_D^3 = 0.2 \gev^3$ and $\rho_{LS}^3 = -0.15 \gev^3$.
Perturbative uncertainties are addressed by varying $\alpha_s$ by $\pm 0.1$ for hadron moments and 
by $\pm 0.04$ for lepton and photon moments around a central value of $\alpha_s = 0.22$. 
Finally, we also vary the quark masses $m_b$ and $m_c$ by 20 \mev\ around central values of 4.6 \gev and 1.18 \gev. 

These uncertainties $\sigma_{\xi_i}$ are propagated into an error on the individual moments $\Delta M$ using Gaussian error propagation
\begin{equation}
(\Delta M)^2  =\sum_{i} (\frac{\partial M}{\partial \xi_i})^2  \sigma_{\xi_i}^2
\end{equation}
where $\xi = (m_b,m_c,\mu_{\pi}^2,\mu_{G}^2,\rho_{D}^3,\rho_{LS}^3,\alpha_s )$.
All these variations are considered uncorrelated for a given moment. 
The theoretical covariance matrix is then constructed by treating these errors as fully correlated for a given moment with different $E_{\rm cut}$ while they are treated as uncorrelated between moments of different order.

For the moments of the photon energy spectrum we include additional theoretical errors related to the applied bias corrections  
as described in Ref.~\cite{Benson:2004sg}. 
We follow the suggestion of the authors and take 30\% of the absolute value of the particular bias as its uncertainty~\cite{kolyapriv}. 
In addition we linearly add half the difference in the moments derived from the two distribution function ans\"atze as given in Ref.~\cite{Benson:2004sg}. 
These additional theoretical errors related to the photon energy moments are considered to be uncorrelated for moments with different $E_{\rm cut}$ and of different order. 

Generally, the chosen approach for the evaluation of the theoretical uncertainties is conservative. 
As a result of this the $\chi^2/N_{\rm dof}$ is very good as will be shown in the next section. 
It is interesting to note though, that a fit that neglects the theoretical errors leads to similar results.
However, we consider the choice of theoretical errors as appropriate since we are trying to consistently extract \Vcb and the heavy quark distribution function parameters from a fit to all moment measurements, and aim to arrive at a reliable error on \Vub when using these results as input.
The conservative approach is also reflected in the fact that we exclude non-integer hadron moments and photon moments above $E_{\rm cut} = 2.0 \gev$ from the fit.

\subsection{Fit Results}

In the following we present the results of a combined fit of the HQEs to all moment measurements listed in Table~\ref{tab:momsum} along with their corresponding theoretical error estimates as defined in Section~\ref{theory_errors}.

 \begin{figure*}[t]
    \begin{center}
\epsfig{file=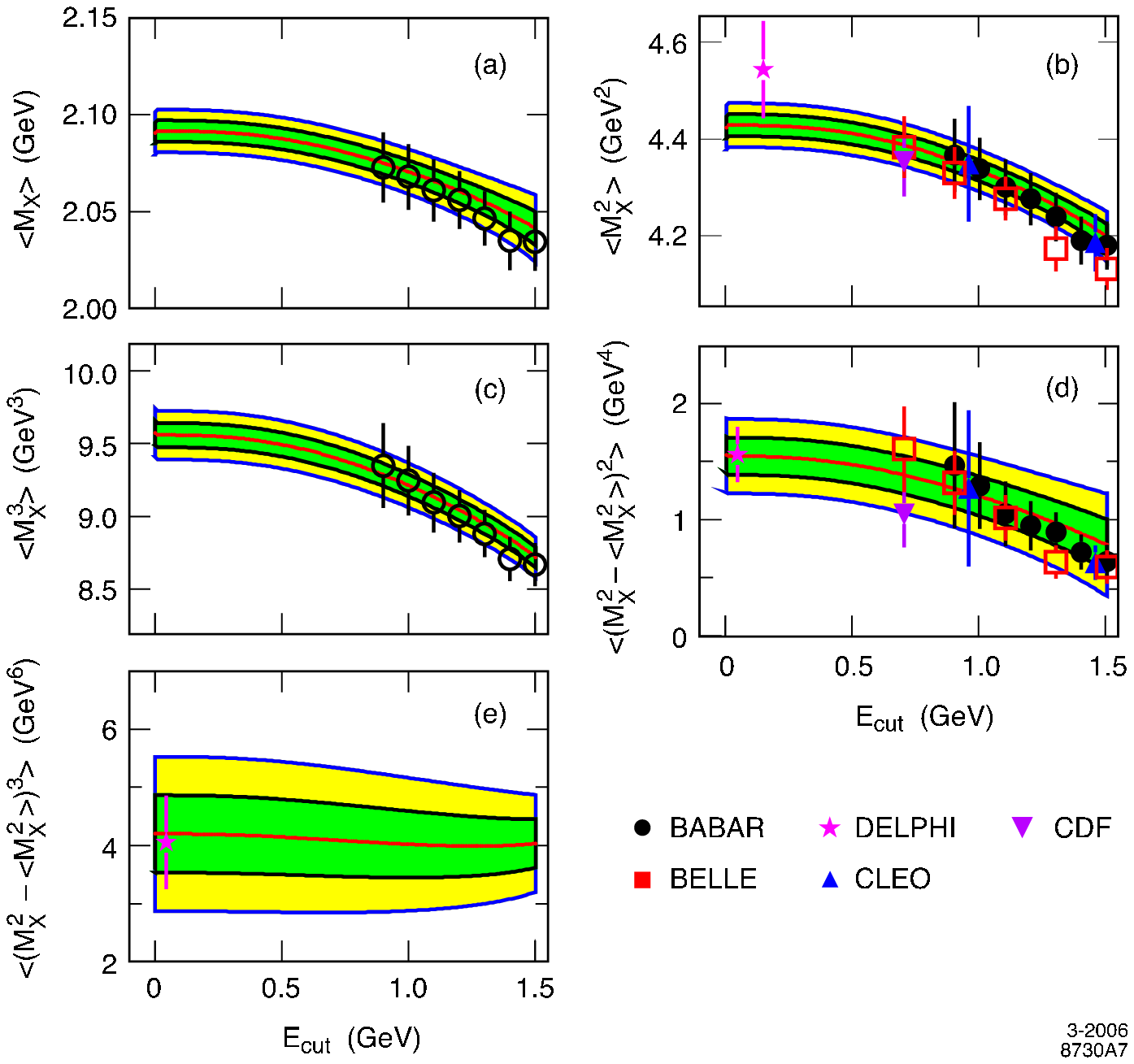}
    \caption{\label{fig:bestfithad_theo}
Comparison of fit predictions and the hadron moment measurements: (a) $\langle M_X\rangle $, (b) $\langle M_X^2\rangle $, (c) $\langle M_X^3\rangle $, (d) $\langle (M_X^2-\langle M_X^2\rangle )^2\rangle $ and  (e) $\langle (M_X^2-\langle M_X^2\rangle )^3\rangle $. The yellow bands represent the total experimental and theoretical fit uncertainty as obtained by converting the fit errors of each individual HQE parameter into an error for the individual moment. The green band indicates the experimental uncertainty only. Solid markers are included in the fit while open markers are only overlaid for comparison. Moment measurements at different $E_{cut}$ are highly correlated.
}
   \end{center}
   \end{figure*}

\begin{figure*}[t]
    \begin{center}

\epsfig{file=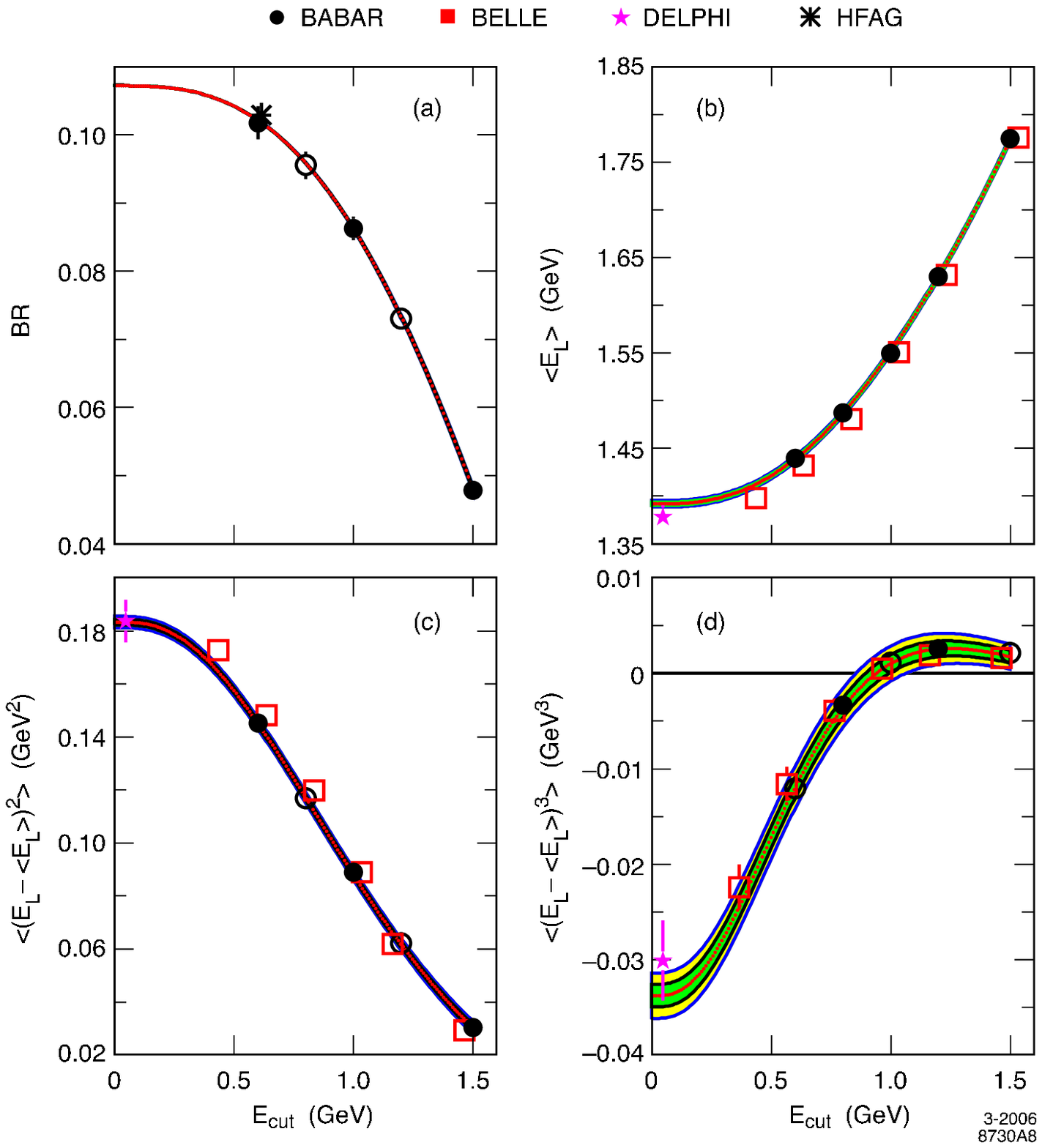}

    \caption{\label{fig:bestfitlep_theo}
Comparison of fit predictions and measurements for the lepton moments: (a) $BR$, (b) $\langle E_L\rangle$, (c) $\langle (E_L-\langle E_L\rangle )^2\rangle $ and (d) $\langle (E_L-\langle E_L\rangle )^3\rangle$. The yellow bands represent the total experimental and theoretical fit uncertainty while the green band indicates the experimental uncertainty only. Solid markers are included in the fit while open markers are only overlaid for comparison. Moment measurements at different $E_{cut}$ are highly correlated.}
   \end{center}
   \end{figure*}

\begin{figure*}[ht!]
    \begin{center}

\epsfig{file=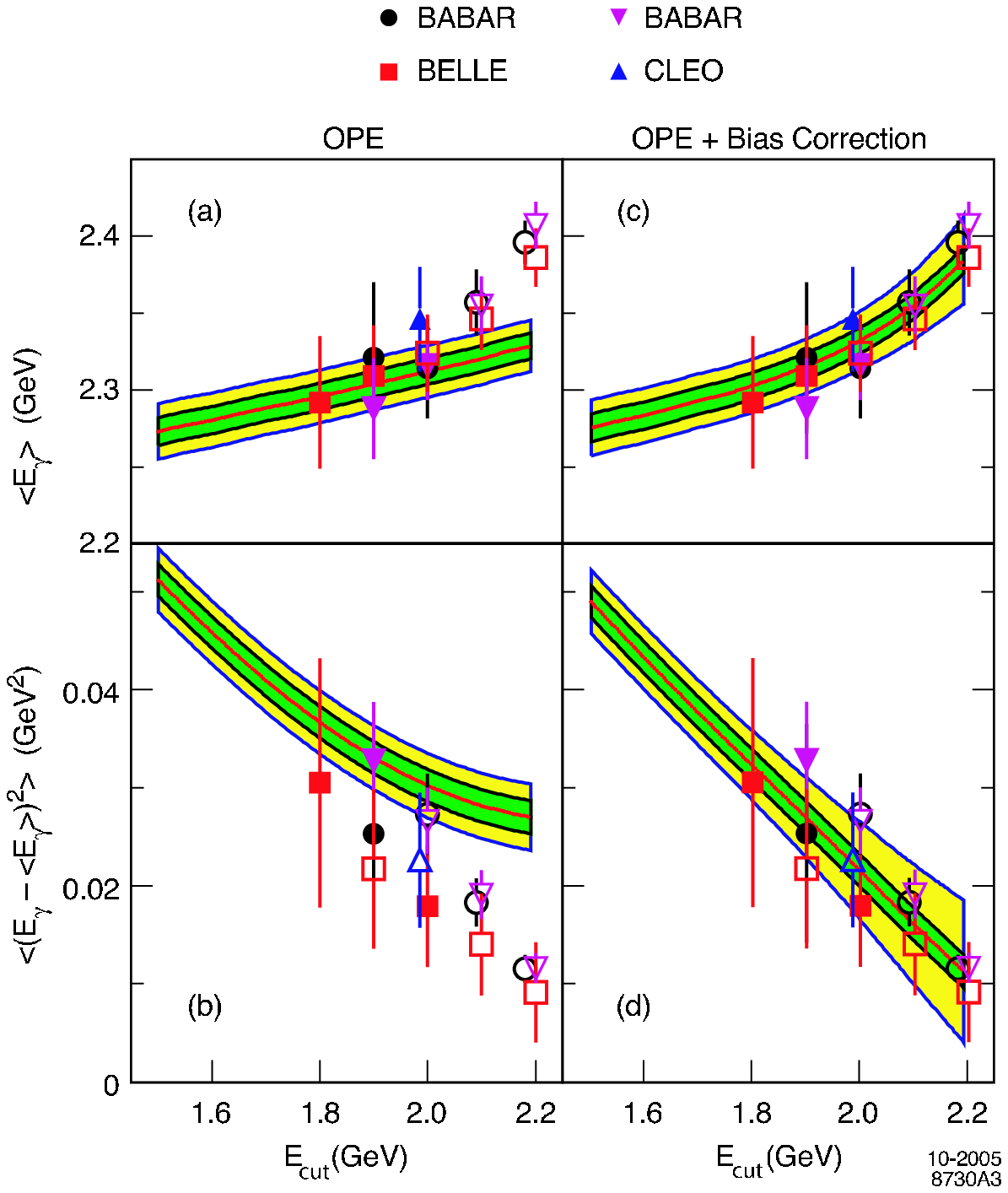}

    \caption{
Comparison of fit predictions and measurements for the photon moments: (a) \meg\ and (b) \vegs.  The bands in the figures on the left show the fit prediction for the pure OPE calculation neglecting effects of the minimal photon energy cut on the OPE part (biases).
The bands in figures (c) and (d) include those bias corrections of Ref.~\cite{Benson:2004sg}.
The yellow bands represent the total experimental and theoretical fit uncertainty while the green band indicates the experimental uncertainty only. Solid markers are included in the fit while open markers are only overlaid for comparison. Moment measurements at different $E_{cut}$ are highly correlated. 
\label{fig:bestfitphot_theo}}
      \end{center}
   \end{figure*}

In order to asses the consistency of the moment measurements from the two different decay processes, \btoclv\ and \btosg\ , we also carry out separate fits to \btoclv\ moments and to photon moments only. However, as the latter are not sensitive to all the heavy quark parameters, all but $m_b$ and $\mu_{\pi}^2$ are fixed to the result obtained from the 
combined fit.\\

A detailed comparison of the HQE predictions obtained from the combined fit and the moment 
measurements is shown in Figures~\ref{fig:bestfithad_theo}-\ref{fig:bestfitphot_theo} for the 
hadron, lepton and photon moments, respectively. The yellow bands represent the total experimental 
and theoretical fit uncertainty as obtained by converting the fit errors for each individual HQE 
parameter into an error for the individual moment. 
The green band indicates the experimental error only.
These figures also show the measurements that are not included in the fit for the reasons 
described in Section~\ref{exp_input}. In particular the non-integer hadron moments \mmx\ and 
\mmxd\ can therefore be directly compared with the corresponding fit prediction. 
It can be seen that all moment measurements agree with each other and that the fit is able to 
describe all the moment measurements of different order and from different $B$ decay distributions.

Figure~\ref{fig:bestfitphot_theo} shows a comparison of the fit prediction for the first and
second moments of the photon energy spectrum for the standard OPE ansatz with the 
$E_{\rm cut}$ dependent bias corrected OPE calculations of Ref.~\cite{Benson:2004sg}.
While it is expected that the pure local OPE approach will break down at higher values of 
$E_{\rm cut}$ this is the first time that the accuracy of the experimental moment 
measurements is sufficient to demonstrate this effect. 
For $E_{\rm cut}$ above 2.0~\gev it is clearly visible that the pure OPE ansatz fails to describe the data.\\

\begin{figure*}[t]
    \begin{center}
\epsfig{file=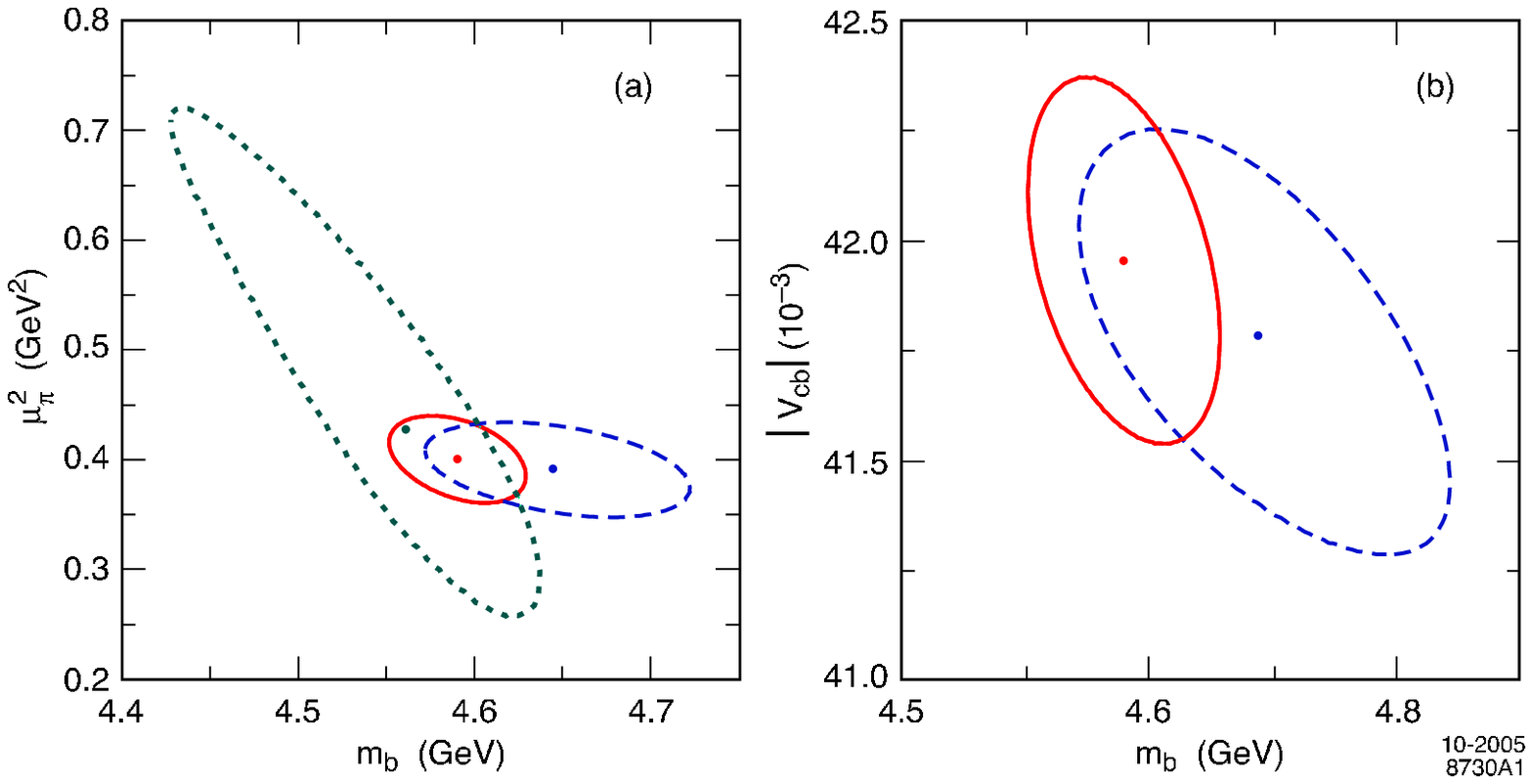}
    \caption{\label{fig:def_result} Comparison of the different fit scenarios. Figure (a) shows the $\Delta \chi^2$ = 1 contour in the ($m_b$,$\mu_{\pi}^2$) plane for the combined fit to all moments (solid red), the fit to hadron and lepton moments only (dashed blue) and the fit to photon moments only (dotted green). 
Figure (b) shows the results for the combined fit (solid red) and the fit to hadron and lepton moments only (dashed blue) in the ($m_b$,\vcb) plane.}
    \end{center}
\end{figure*}

The fit results are summarised in Table~\ref{tab:def_result} where the separation of the errors into experimental and theoretical contributions was obtained from toy Monte Carlo experiments.
The results are in good agreement with earlier 
determinations~\cite{BABAR,DELPHI,Bauer:2004ve} but have improved accuracy.
A comparison of results from the combined fit with those obtained from fits to \btoclv\ and \btosg\ moments only can be 
found in Figure~\ref{fig:def_result} where the $\Delta \chi^2 =1$ contours for the fit results are shown in the ($m_b,\vcb$) and 
($m_b,\mu_{\pi}^2$) planes. It can be seen that the inclusion of the photon energy moments adds additional sensitivity 
to the b-quark mass $m_b$.

\begin{table*}[t]
\caption[]{\label{tab:def_result} Results for the combined fit to all moments with experimental and theoretical uncertainties.
For \Vcb\ we add an additional theoretical error stemming from the uncertainty in the expansion for $\Gamma_{\rm SL}$ of 1.4\%. Below the fit results the correlation matrix is shown.}
\begin{center}
\normalsize
\begin{tabular}{c|cccccccc}\hline\hline
\verysmallrule Combined& \multicolumn{8}{c}{OPE FIT RESULT: $\chi^2/N_{dof}$ =19.3/44} \\
   Fit        &$|V_{cb}|$ $\times 10^{-3}$ &$m_b$ ($\gev$) &$m_c$ ($\gev$) &$\mu_{\pi}^2$ ($\gev^2$) &   $\rho_D^3$ ($\gev^3$) &  $\mu_{G}^2$ ($\gev^2$) &$\rho_{LS}^3$ ($\gev^3$) & $BR_{c\ell\bar{\nu}}$ ($\%$)    \\ \hline 
RESULT       &41.96 &4.590 &1.142 &0.401 &0.174 &0.297 &-0.183 &10.71    \\     
$\Delta$ exp & 0.23 &0.025 &0.037 &0.019 &0.009 &0.024 & 0.054 & 0.10 \\ 
$\Delta$ HQE & 0.35 &0.030 &0.045 &0.035 &0.022 &0.046 & 0.071 & 0.08 \\ 
$\Delta$ \gsl& 0.59 &      &      &      &      &      &       &      \\ \hline 

$|V_{cb}|$    &  1.000 & -0.399 & -0.220 &  0.405 &  0.267 & -0.305 &  0.056 &  0.700 \\
$m_b$         &        &  1.000 &  0.951 & -0.387 & -0.189 &  0.074 & -0.223 &  0.098 \\
$m_c$         &        &        &  1.000 & -0.408 & -0.246 & -0.329 & -0.124 &  0.143 \\
$\mu_{\pi}^2$ &        &        &        &  1.000 &  0.685 &  0.257 & -0.008 &  0.122 \\
$\rho_D^3$    &        &        &        &        &  1.000 & -0.050 & -0.479 & -0.055 \\
$\mu_{G}^2$   &        &        &        &        &        &  1.000 & -0.035 &  0.046 \\
$\rho_{LS}^3$ &        &        &        &        &        &        &  1.000 & -0.052 \\
$BR_{c\ell\bar{\nu}}$  &        &        &        &        &        &        &        &  1.000 \\ \hline\hline 
\end{tabular}
\end{center}
\end{table*}

Including the Belle measurements of the hadron mass and lepton energy moments with only their 
statistical correlations leads to very similar results with only small improvements in the errors for 
the heavy quark parameters. This is a consequence of the fit errors being dominated by the 
theoretical uncertainties as can be seen from Table~\ref{tab:def_result}. For a future average based on
the full covariance matrix it will be necessary to consider the strong correlation of the systematic 
errors of the lepton moments between experiments. 
In particular, a consistent background modeling and subtraction technique will be required.

To ensure the stability of the fit procedure several cross checks have been carried out. For instance, the combined fit 
has been repeated without applying the theoretical constraints on $\mu_G^2$ and $\rho_{LS}^3$. We also repeated the fit excluding hadron moments with units $\gev^6$ and lepton moments with units $\gev^3$ as these moments are believed to have large theoretical uncertainties (as can bee seen from Figures~\ref{fig:bestfithad_theo} and~\ref{fig:bestfitlep_theo}). In addition, photon moments with $E_{\rm cut} > 1.8 \gev$ were excluded as here the bias corrections become noticeable.
Finally a fit neglecting all theoretical errors was performed, i.e. only the experimental 
covariance matrix was used.  
All these results agree well with each other and any variations are fully covered by the 
theoretical error estimates.
In addition the scale dependence of the expressions for the moments was studied but was found 
to be small compared to the assigned theoretical uncertainties.\\

In addition to the above we extract the difference in the quark masses as
\begin{eqnarray}
 m_b - m_c = 3.446 \pm 0.025 \gev \nonumber \,.  
\end{eqnarray}

Comparing the extracted values of the quark masses $m_b$ and $m_c$
with other determinations is often convenient in the commonly used 
$\overline {\rm MS}$  scheme. The translation between the kinetic and
$\overline {\rm MS}$ masses to two loop accuracy and including 
the BLM part of the
$\alpha_s^3$ corrections was given in Ref.~\cite{Kolya}. This leads to
\begin{eqnarray}
\overline{m_b}(\overline{m_b})=4.20\pm 0.04\,\gev \nonumber \\
\overline{m_c}(\overline{m_c})=1.24\pm 0.07\,\gev \nonumber  
\end{eqnarray}
These results agree well with the determination in the 1S scheme~\cite{Bauer:2004ve,Hoang:2005zw} 
and recent unquenched lattice calculations~\cite{Gray:2005ur,DiRenzo:2004xn,Nobes:2005dz}. 
However, it has been accepted among theorists that the
normalization scale of around 1.2 \gev\ in the $\overline {\rm MS}$ scheme 
may be too low for a precision evaluation of masses, and higher-order
perturbative corrections in $\overline{m_c}(\overline{m_c})$ 
are too significant. As a result, an
additional uncertainty in $\overline{m_c}(\overline{m_c})$ of at least
$50\,\mev$ may have to be added associated with the definition of
$\overline{m_c}(\overline{m_c})$ itself. A larger normalization scale
for the $\overline {\rm MS}$ masses is generally used.  To address
this we give here the value of $m_c$ normalized at a safer momentum
scale $2.5\,\gev$ as was advocated recently:
\begin{equation}
\overline{m_c}(2.5\gev)=1.072\pm 0.06 \,\gev\,. \nonumber
\end{equation}
The theoretical uncertainty in this translation is small. It may also
be convenient to have the ratio of the charm and the beauty quark masses 
in the $\overline {\rm MS}$ scheme which is normalization-scale
independent:
\begin{equation}
\frac{\overline{m_c}(\mu)}{\overline{m_b}(\mu)}=0.235\pm 0.012 \,.  \nonumber
\end{equation}
The uncertainty in this ratio is dominated by the fit error on $m_c$.

\clearpage

\section{Translation of fit results into other schemes}

We translate the results for $m_b$ and $\mu_{\pi}^2$ in the kinetic scheme to heavy quark 
distribution function parameters in other schemes so that they can be used for the 
extraction of \Vub.
The translation is done by predicting the first and second moment of the photon energy 
spectrum above $E_{\rm cut} = 1.6 \gev$ based on the heavy quark parameters from 
Table~\ref{tab:def_result} and using the calculations of Ref.~\cite{Benson:2004sg}.

The experimental and theoretical uncertainties in the fitted parameters as well as their 
correlations are propagated into the errors on the moments as described in 
Section~\ref{theory_errors}.
The minimum photon energy of 1.6 \gev is chosen such as to be insensitive to the distribution 
function itself. At this threshold the local OPE calculation is applicable as the hardness 
${\cal Q} = m_B - 2 E_{\rm cut}$ of the process is sufficiently high such that cut-induced 
perturbative and non-perturbative corrections or biases are negligible.
The predicted moments are given in Table~\ref{tab:egamma-pred}.

As the moments are physical observables which are scheme independent they can be used to 
extract the corresponding heavy quark distribution function parameters in other schemes.
For this translation, grids for the first and second moments of the photon energy spectrum are generated as a function of the two parameters ($\bar{\Lambda}$,$\lambda_1$) for Kagan-Neubert~\cite{Kagan} and ($m_{b~SF}$,$\mu_{\pi~SF}^2$) for the Shape Function~\cite{Bosch:2004th} 
scheme. 
A $\chi^2$ is calculated for every set of parameters $\mu = (\meg({m_b},{\mu_{\pi}^2}),\vegs({m_b},{\mu_{\pi}^2}))$ as
\begin{equation}
 \chi^2 = \sum_{i,j=1,2} (y_i-\mu_i) \, V_{ij}^{-1} \, (y_j-\mu_j) \,\, \mbox{with} \,\, V_{ij}= \sigma_i \sigma_j \rho_{ij} 
\end{equation}
where the $y_i$ are the predicted moments with their errors $\sigma_i$ and $\rho_{ij}$ is the correlation between them.

From the minimum value $\chi^2_{min}$ we obtain the central values for the parameters in the other
schemes and determine the $\Delta \chi^2 = 1 $ contour with respect to $\chi^2_{min}$.

\subsection{Kagan-Neubert scheme}
\begin{figure*}[t]
    \begin{center}
\epsfig{file=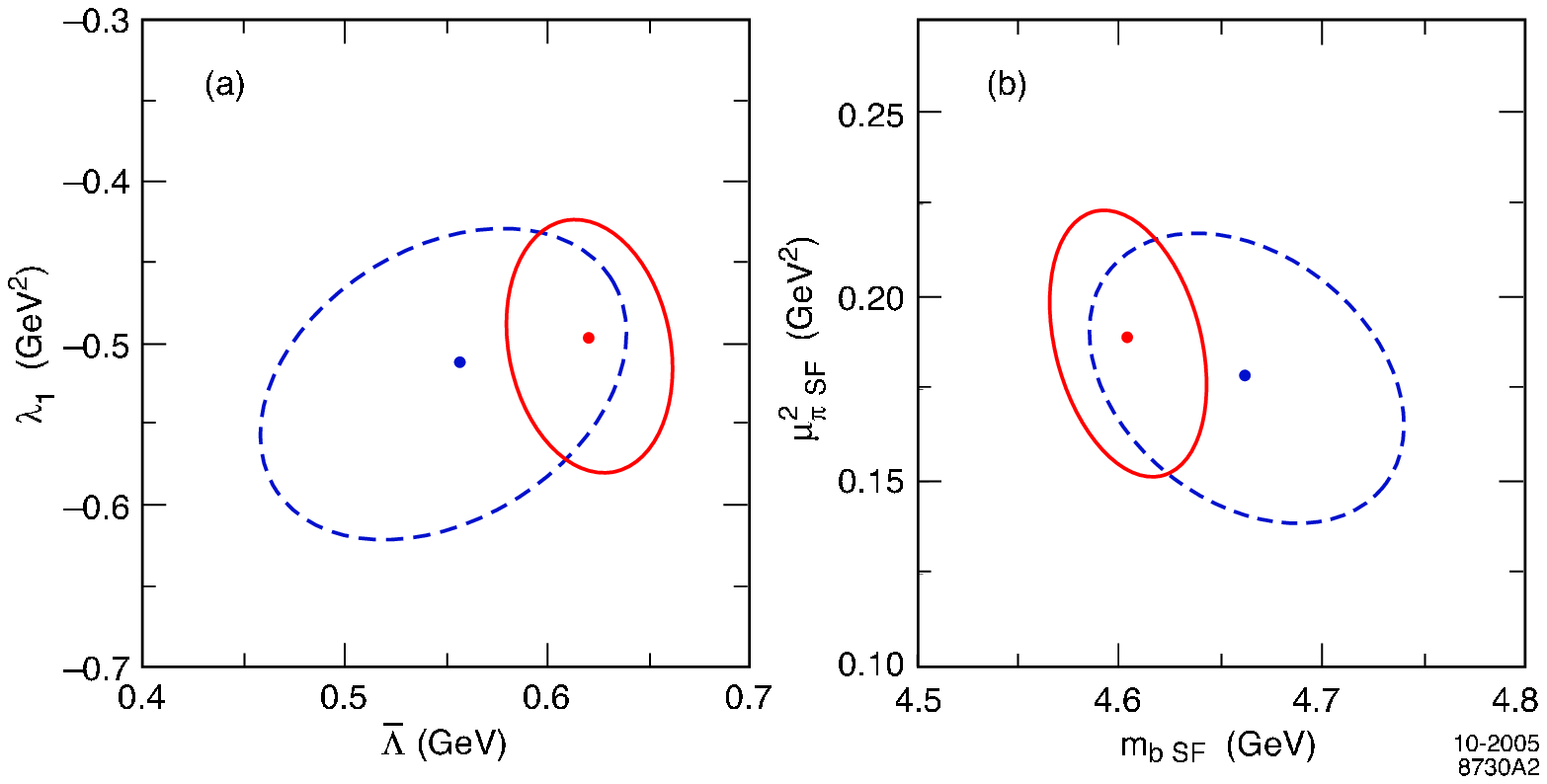}
    \caption{\label{fig:kn-comp} Translation of fit results in the kinetic scheme to Kagan-Neubert (a) and the Shape Function scheme (b) via predicted photon moments. Figure (a) shows the results for the shape function parameters in the ($\bar{\Lambda},\lambda_1$) plane from the combined fit to all moments (solid red) and the fit to hadron and lepton moments only (dashed blue). Figure (b) shows the corresponding fit results in the ($m_{b~SF},\mu_{\pi~SF}^2$) plane.}
    \end{center}
\end{figure*}

In order to derive shape function parameters from the predicted moments in the Kagan-Neubert scheme~\cite{Kagan} a grid of moments was generated for varying values of $\bar{\Lambda}$ and $\lambda_1$. This was obtained using the Kagan-Neubert $B \ra X_s \gamma$ generator with the exponential Shape Function ansatz as implemented in the Babar Monte Carlo generator (EvtGen~\cite{Lange:2001uf}).
The results of this translation are shown in Table~\ref{tab:kn-comp} and Figure~\ref{fig:kn-comp}.

\begin{table}[]
\begin{center}
\normalsize
\caption[]{\label{tab:egamma-pred} First and second moment of the photon spectrum predicted for $E_{\rm cut} = 1.6 \gev$ on the basis of the fit results for the HQE parameters. 
}
\begin{tabular}{cccc}\hline
\vvsmallrule $E_{\rm cut} (\gev)$ & $\meg (\gev)$ & $ \vegs (\gev^2)$  & $\rho$  \\  
\vvsmallrule   1.6     & 2.284 $\pm$ 0.018 & 0.0428 $\pm$ 0.0032 & -0.03  \\ \hline
\end{tabular}
\end{center}
\end{table}

\begin{table}[]
\begin{center}
\normalsize
\caption[]{\label{tab:kn-comp} 
Comparison of heavy quark distribution function parameters in the kinetic, Kagan-Neubert and Shape Function scheme together with their correlation $\rho$.}
\begin{tabular}{ccc}\hline
 \multicolumn{3}{c}{\vvsmallrule Kinetic Scheme}\\
      $m_b (\gev)$     & $\mu_{\pi}^2 (\gev^2)$  & $\rho$  \\
      $4.590 \pm0.039$  & $0.401 \pm 0.040$      & $-0.39$\\ \hline 
 \multicolumn{3}{c}{\vvsmallrule Kagan-Neubert Scheme}\\
      $\bar{\Lambda} (\gev)$ & $\lambda_1(\gev^2)$                  & $\rho$  \\
      $0.621 \pm0.041$       & $-0.497~_{ -~  0.086}^{ + ~ 0.072}$  & $-0.17$\\ \hline
 \multicolumn{3}{c}{\vvsmallrule Shape Function Scheme}\\   
      $m_{b~SF} (\gev)$ &  $\mu_{\pi~SF}^2(\gev^2)$ & $\rho$\\ 
      $4.604 \pm 0.038$ &  $0.189 \pm 0.038$        & $-0.23$\\ \hline 
\end{tabular}
\end{center}
\end{table}

\subsection{Shape-Function scheme}
For the translation into the Shape-Function scheme~\cite{Bosch:2004th,Neubert:2004sp} we use a grid of 
moments obtained with a {\it Mathematica} notebook based on Ref.~\cite{Lange:2005yw,Neubert:2004cu,Bosch:2004cb,Neubert:2004dd} that was provided to us by the authors. 
In this calculation the moments are determined from a spectrum that is obtained by convoluting a shape function with 
a perturbative kernel with next-to-leading order accuracy, where we use the exponential form for the shape function 
given in Ref.~\cite{Lange:2005yw}.
This calculation is conceptually similar to the one for $B \ra X_u \ell \bar{\nu}$ decays also presented in Ref.~\cite{Lange:2005yw}
which at present is 
used for the extraction of \Vub\ by several experiments. It therefore allows for a consistent 
determination of the shape function parameters for both, \btosg\ and $B \ra X_u \ell \bar{\nu}$ decays.
The numerical results for the shape function parameters are shown in Table~\ref{tab:kn-comp} 
and the $\Delta \chi^2 = 1$ contours are displayed in Figure~\ref{fig:kn-comp}.

\section{Applications for improved heavy quark parameters}

\subsection{Improved OPE Expression for \Vub}

The results in the kinetic scheme for $m_b$, $\mu_{\pi}^2$, $\mu_G^2$ and $\rho_D^3$ have been used to give an updated expression for the standard local OPE formula for \Vub of Ref.~\cite{Uraltsev:1999rr,kolyapriv}: 
\begin{eqnarray}
\Vub &=& 4.268 \cdot 10^{-3} \cdot \sqrt{\frac{BR(B\to X_u \ell \bar{\nu})}{0.002}\frac{1.61 ps}{\tau_B}} \nonumber \\
& &\times (1 \pm 0.012_{\rm QCD} \pm 0.022_{\rm HQE})  \,.
\end{eqnarray}
The error labeled `QCD' includes perturbative uncertainties and those from weak annihilation.
However, in contrast to Ref.~\cite{Uraltsev:1999rr} we also include explicitely the $1/m_b^3$ contribution from the $\rho_D^3$ term which results in an improved `QCD' error.
The `HQE' related uncertainty stems from the errors on $m_b$, $\mu_{\pi}^2$, $\mu_G^2$ and $\rho_D^3$ 
and takes correlations between the parameters into account (see Table~\ref{tab:def_result}).

\subsection{Extrapolation Factors for Measured \btosg\ Branching Fraction}

\begin{figure*}[t]
    \begin{center}
\epsfig{file=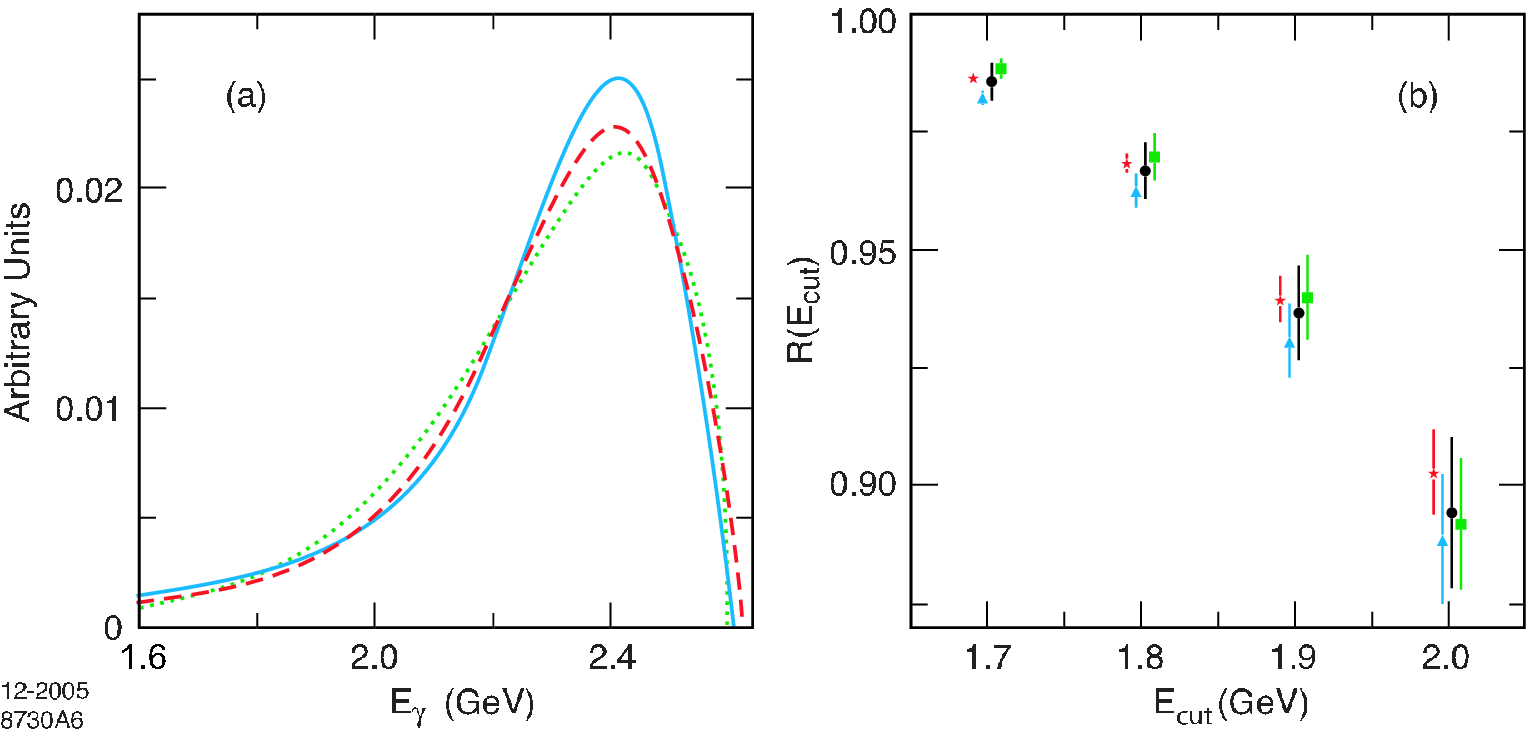}
    \caption{\label{fig:spectra} Figure (a) shows the photon energy spectra corresponding to the fitted heavy quark distribution parameters in the kinetic scheme (dashed red), Shape Function scheme (solid blue) and Kagan-Neubert scheme (dotted green). Figure (b) shows the corresponding extrapolation factors $R(\ecut)$ for varying \ecut\ for the kinetic scheme (red stars), Shape Function scheme (blue triangles) and Kagan-Neubert scheme (green squares) together with our average (black circles).}
    \end{center}
\end{figure*}

The measurement of the \btosg\ branching fraction is experimentally very challenging and has 
only been achieved for photon energies above $\ecut = 1.8 - 2.0 \gev$. On the contrary, theoretical calculations predict the \btosg\ branching fraction at much lower values of \ecut\ in order to avoid any dependence on the heavy quark distribution function. 
It is therefore customary to extrapolate measured branching fractions down to a value of 1.6 \gev where they can be compared to the theoretical calculations~\cite{Gambino:2001ew,Buras:2002tp}. 
Based on the heavy quark distribution function parameters in Table~\ref{tab:kn-comp} and the corresponding spectra we calculated a consistent set of extrapolation factors 
\begin{equation}
R(E_{\rm cut}) = \frac{BR(\btosg)_{E_{\rm cut}}}{BR(\btosg)_{1.6 \gev}}
\end{equation}
for the kinetic, Kagan-Neubert and Shape-Function scheme. The results are summarised in Table~\ref{tab:bsg-extra} and Figure~\ref{fig:spectra}. 
The error was determined as the largest deviation from the central value obtained from a scan around the ellipses in Figures~\ref{fig:def_result} and~\ref{fig:kn-comp}, where positive and negative errors were of comparable size. 
The results have been averaged where the total error was determined by combining the largest error from 
the scan of the error ellipses with half the maximum difference between any two models in quadrature.
Figure~\ref{fig:spectra} also shows the spectra corresponding to the central values of Table~\ref{tab:kn-comp} or equivalently to the predicted photon energy moments of Table~\ref{tab:egamma-pred} in the three schemes.

\begin{table*}[htbp]
\begin{center}
\normalsize
\caption[]{\label{tab:bsg-extra} Extrapolation factors $R(E_{\rm cut})$ for $BR(\btosg)$. }
\begin{tabular}{ccccc}\hline
\vvsmallrule& Kinetic &Kagan-Neubert &Shape Function& Average \\
& Scheme & Scheme &Scheme &\\
\vvsmallrule $E_{\rm cut} (\gev)$ & $R(E_{\rm cut})$ &  $R(E_{\rm cut})$ & $R(E_{\rm cut})$& $R(E_{\rm cut})$ \\  \hline
              1.7     & 0.986 $\pm$ 0.001 & 0.988 $\pm$ 0.002 & 0.982 $\pm$ 0.002 & 0.985 $\pm$ 0.004\\ 
              1.8     & 0.968 $\pm$ 0.002 & 0.970 $\pm$ 0.005 & 0.962 $\pm$ 0.004 & 0.967 $\pm$ 0.006\\ 
              1.9     & 0.939 $\pm$ 0.005 & 0.940 $\pm$ 0.009 & 0.930 $\pm$ 0.008 & 0.936 $\pm$ 0.010\\ 
              2.0     & 0.903 $\pm$ 0.009 & 0.892 $\pm$ 0.014 & 0.888 $\pm$ 0.014 & 0.894 $\pm$ 0.016\\ \hline
\end{tabular}
\end{center}
\end{table*}

\section{Conclusion}

We have performed a fit to moments measurements from \btoclv\ and \btosg\ decays using calculations in the kinetic scheme~\cite{KolyaPaolo,Kolya,Benson:2004sg}. The fit uses all currently available moment measurements from the \babar\, Belle, CDF, CLEO and DELPHI experiments that are
publicly available with their corresponding correlation matrices. 
We find that all the moment measurements of different order and from different inclusive $B$ 
decays can be described by the fit result which is an important test of the consistency
of this theoretical framework.
We have extracted values for the CKM matrix element \Vcb, the quark masses $m_b$ and $m_c$, and the kinetic expectation value $\mu_{\pi}^2$ of
\begin{eqnarray}
\Vcb &=&( 41.96 \pm 0.23_{\rm exp} \pm 0.35_{\rm HQE} \pm 0.59_{\gsl} ) \times 10^{-3} \nonumber \\ 
m_b &=& 4.590 \pm 0.025_{\rm exp} \pm 0.030_{\rm HQE} \gev\nonumber \\ 
m_c &=& 1.142 \pm 0.037_{\rm exp} \pm 0.045_{\rm HQE} \gev\nonumber \\ 
\mu_{\pi}^2 &=& 0.401 \pm 0.019_{\rm exp} \pm 0.035_{\rm HQE} \gev^2\nonumber 
\end{eqnarray}
where the first error includes statistical and systematic experimental uncertainties and the second the theoretical uncertainties from the HQEs.\\
As can be seen, the error on \Vcb which is below 2\% is dominated by theorectical uncertainties.
Any further improvements will require additional work on the accuracy for the expression of \gsl, in particular
on perturbative corrections to the Wilson coefficients of the the chromomagnetic and Darwin operators.
Similar observations can be made for $m_b$, which is determined with better than 1\% accuracy. 
However, the extraction of these quantities at the percent level represents in itself a remarkable test 
and success of the QCD-based calculations.

The values for $m_b$ and $\mu_{\pi}^2$ have been translated into the Kagan-Neubert Scheme where we obtain 
following values for the Shape Function parameters:
\begin{eqnarray}
\bar{\Lambda}   &=& 0.621 \pm 0.041 \gev\nonumber \\ 
\lambda_1 &=& -0.497^{+0.072}_{-0.086} \gev^2\nonumber  \quad .
\end{eqnarray}
Similarly, we obtain 
\begin{eqnarray}
m_{b~SF} &=& 4.604 \pm 0.038 \gev\nonumber \\ 
\mu_{\pi~SF}^2 &=& 0.189 \pm 0.038 \gev^2\nonumber 
\end{eqnarray}
in the Shape Function scheme.
As these parameters are critical for the extraction of \Vub, their reduced uncertainty 
will enable measurements of \Vub\ at the 5\% level.
This, together with \Vcb, will provide for a competitive measurement of the side of the Unitarity Triangle
opposite the angle $\beta$, and give further insights into the extent of CP violation in tree processes.\\

\section{Acknowledgements}

This work has greatly benefited from many interactions and exchanges with N. Uraltsev, P. Gambino and I. Bigi
concerning the calculations in the kinetic scheme. 
The authors would also like to thank M. Neubert for many discussions regarding the calculations for 
\btosg\ decays in the shape function scheme. 
Furthermore, we would like to thank our colleagues of the \babar\ Collaboration, in particular F. di Lodovico 
and V. L\"uth, and our colleagues from HFAG for their valuable input to this analysis.
H. F. is supported by a PPARC Postdoctoral Fellowship.


\begin{thebibliography}{46}
\expandafter\ifx\csname natexlab\endcsname\relax\def\natexlab#1{#1}\fi
\expandafter\ifx\csname bibnamefont\endcsname\relax
  \def\bibnamefont#1{#1}\fi
\expandafter\ifx\csname bibfnamefont\endcsname\relax
  \def\bibfnamefont#1{#1}\fi
\expandafter\ifx\csname citenamefont\endcsname\relax
  \def\citenamefont#1{#1}\fi
\expandafter\ifx\csname url\endcsname\relax
  \def\url#1{\texttt{#1}}\fi
\expandafter\ifx\csname urlprefix\endcsname\relax\def\urlprefix{URL }\fi
\providecommand{\bibinfo}[2]{#2}
\providecommand{\eprint}[2][]{\url{#2}}

\bibitem[{\citenamefont{Falk and Luke}(1998)}]{Falk:1997jq}
\bibinfo{author}{\bibfnamefont{A.~F.} \bibnamefont{Falk}} \bibnamefont{and}
  \bibinfo{author}{\bibfnamefont{M.~E.} \bibnamefont{Luke}},
  \bibinfo{journal}{Phys. Rev.} \textbf{\bibinfo{volume}{D57}},
  \bibinfo{pages}{424} (\bibinfo{year}{1998}), \eprint{hep-ph/9708327}.

\bibitem[{\citenamefont{Gambino and Uraltsev}(2004)}]{KolyaPaolo}
\bibinfo{author}{\bibfnamefont{P.}~\bibnamefont{Gambino}} \bibnamefont{and}
  \bibinfo{author}{\bibfnamefont{N.}~\bibnamefont{Uraltsev}},
  \bibinfo{journal}{Eur. Phys. J.} \textbf{\bibinfo{volume}{C34}},
  \bibinfo{pages}{181} (\bibinfo{year}{2004}), \eprint{hep-ph/0401063}.

\bibitem[{\citenamefont{Benson et~al.}(2003)\citenamefont{Benson, Bigi, Mannel,
  and Uraltsev}}]{Kolya}
\bibinfo{author}{\bibfnamefont{D.}~\bibnamefont{Benson}},
  \bibinfo{author}{\bibfnamefont{I.~I.} \bibnamefont{Bigi}},
  \bibinfo{author}{\bibfnamefont{T.}~\bibnamefont{Mannel}}, \bibnamefont{and}
  \bibinfo{author}{\bibfnamefont{N.}~\bibnamefont{Uraltsev}},
  \bibinfo{journal}{Nucl. Phys.} \textbf{\bibinfo{volume}{B665}},
  \bibinfo{pages}{367} (\bibinfo{year}{2003}), \eprint{hep-ph/0302262}.

\bibitem[{\citenamefont{Benson et~al.}(2005)\citenamefont{Benson, Bigi, and
  Uraltsev}}]{Benson:2004sg}
\bibinfo{author}{\bibfnamefont{D.}~\bibnamefont{Benson}},
  \bibinfo{author}{\bibfnamefont{I.~I.} \bibnamefont{Bigi}}, \bibnamefont{and}
  \bibinfo{author}{\bibfnamefont{N.}~\bibnamefont{Uraltsev}},
  \bibinfo{journal}{Nucl. Phys.} \textbf{\bibinfo{volume}{B710}},
  \bibinfo{pages}{371} (\bibinfo{year}{2005}), \eprint{hep-ph/0410080}.

\bibitem[{\citenamefont{Bauer et~al.}(2003)\citenamefont{Bauer, Ligeti, Luke,
  and Manohar}}]{Bauer:2002sh}
\bibinfo{author}{\bibfnamefont{C.~W.} \bibnamefont{Bauer}},
  \bibinfo{author}{\bibfnamefont{Z.}~\bibnamefont{Ligeti}},
  \bibinfo{author}{\bibfnamefont{M.}~\bibnamefont{Luke}}, \bibnamefont{and}
  \bibinfo{author}{\bibfnamefont{A.~V.} \bibnamefont{Manohar}},
  \bibinfo{journal}{Phys. Rev.} \textbf{\bibinfo{volume}{D67}},
  \bibinfo{pages}{054012} (\bibinfo{year}{2003}), \eprint{hep-ph/0210027}.

\bibitem[{\citenamefont{Uraltsev}(2005{\natexlab{a}})}]{Uraltsev:2004in}
\bibinfo{author}{\bibfnamefont{N.}~\bibnamefont{Uraltsev}},
  \bibinfo{journal}{Int. J. Mod. Phys.} \textbf{\bibinfo{volume}{A20}},
  \bibinfo{pages}{2099} (\bibinfo{year}{2005}{\natexlab{a}}),
  \eprint{hep-ph/0403166}.

\bibitem[{\citenamefont{Aubert et~al.}(2004{\natexlab{a}})}]{BABAR}
\bibinfo{author}{\bibfnamefont{B.}~\bibnamefont{Aubert}} \bibnamefont{et~al.}
  (\bibinfo{collaboration}{\babar\ Collaboration}), \bibinfo{journal}{Phys.
  Rev. Lett.} \textbf{\bibinfo{volume}{93}}, \bibinfo{pages}{011803}
  (\bibinfo{year}{2004}{\natexlab{a}}), \eprint{hep-ex/0404017}.

\bibitem[{\citenamefont{Bauer et~al.}(2004)\citenamefont{Bauer, Ligeti, Luke,
  Manohar, and Trott}}]{Bauer:2004ve}
\bibinfo{author}{\bibfnamefont{C.~W.} \bibnamefont{Bauer}},
  \bibinfo{author}{\bibfnamefont{Z.}~\bibnamefont{Ligeti}},
  \bibinfo{author}{\bibfnamefont{M.}~\bibnamefont{Luke}},
  \bibinfo{author}{\bibfnamefont{A.~V.} \bibnamefont{Manohar}},
  \bibnamefont{and} \bibinfo{author}{\bibfnamefont{M.}~\bibnamefont{Trott}},
  \bibinfo{journal}{Phys. Rev.} \textbf{\bibinfo{volume}{D70}},
  \bibinfo{pages}{094017} (\bibinfo{year}{2004}), \eprint{hep-ph/0408002}.

\bibitem[{\citenamefont{Abdallah et~al.}(2006)}]{DELPHI}
\bibinfo{author}{\bibfnamefont{J.}~\bibnamefont{Abdallah}} \bibnamefont{et~al.}
  (\bibinfo{collaboration}{DELPHI}), \bibinfo{journal}{Eur. Phys. J.}
  \textbf{\bibinfo{volume}{C45}}, \bibinfo{pages}{35} (\bibinfo{year}{2006}),
  \eprint{hep-ex/0510024}.

\bibitem[{\citenamefont{Bigi et~al.}(1997)\citenamefont{Bigi, Shifman,
  Uraltsev, and Vainshtein}}]{scale}
\bibinfo{author}{\bibfnamefont{I.~I.~Y.} \bibnamefont{Bigi}},
  \bibinfo{author}{\bibfnamefont{M.~A.} \bibnamefont{Shifman}},
  \bibinfo{author}{\bibfnamefont{N.}~\bibnamefont{Uraltsev}}, \bibnamefont{and}
  \bibinfo{author}{\bibfnamefont{A.~I.} \bibnamefont{Vainshtein}},
  \bibinfo{journal}{Phys. Rev.} \textbf{\bibinfo{volume}{D56}},
  \bibinfo{pages}{4017} (\bibinfo{year}{1997}), \eprint{hep-ph/9704245}.

\bibitem[{\citenamefont{Melnikov and Mitov}(2005)}]{Melnikov:2005bx}
\bibinfo{author}{\bibfnamefont{K.}~\bibnamefont{Melnikov}} \bibnamefont{and}
  \bibinfo{author}{\bibfnamefont{A.}~\bibnamefont{Mitov}},
  \bibinfo{journal}{Phys. Lett.} \textbf{\bibinfo{volume}{B620}},
  \bibinfo{pages}{69} (\bibinfo{year}{2005}), \eprint{hep-ph/0505097}.

\bibitem[{\citenamefont{Eidelman et~al.}(2004)}]{PDG2005}
\bibinfo{author}{\bibfnamefont{S.}~\bibnamefont{Eidelman}} \bibnamefont{et~al.}
  (\bibinfo{collaboration}{Particle Data Group}), \bibinfo{journal}{Phys.
  Lett.} \textbf{\bibinfo{volume}{B592}}, \bibinfo{pages}{1}
  (\bibinfo{year}{2004}), \bibinfo{note}{and 2005 partial update for edition
  2006}.

\bibitem[{\citenamefont{Bigi et~al.}(2005)\citenamefont{Bigi, Uraltsev, and
  Zwicky}}]{Bigi:2005bh}
\bibinfo{author}{\bibfnamefont{I.~I.} \bibnamefont{Bigi}},
  \bibinfo{author}{\bibfnamefont{N.}~\bibnamefont{Uraltsev}}, \bibnamefont{and}
  \bibinfo{author}{\bibfnamefont{R.}~\bibnamefont{Zwicky}}
  (\bibinfo{year}{2005}), \eprint{hep-ph/0511158}.

\bibitem[{\citenamefont{Aubert et~al.}(2004{\natexlab{b}})}]{BABARHAD}
\bibinfo{author}{\bibfnamefont{B.}~\bibnamefont{Aubert}} \bibnamefont{et~al.}
  (\bibinfo{collaboration}{\babar\ Collaboration}), \bibinfo{journal}{Phys.
  Rev.} \textbf{\bibinfo{volume}{D69}}, \bibinfo{pages}{111103}
  (\bibinfo{year}{2004}{\natexlab{b}}), \eprint{hep-ex/0403031}.

\bibitem[{\citenamefont{Aubert et~al.}(2004{\natexlab{c}})}]{BABARLEP}
\bibinfo{author}{\bibfnamefont{B.}~\bibnamefont{Aubert}} \bibnamefont{et~al.}
  (\bibinfo{collaboration}{\babar\ Collaboration}), \bibinfo{journal}{Phys.
  Rev.} \textbf{\bibinfo{volume}{D69}}, \bibinfo{pages}{111104}
  (\bibinfo{year}{2004}{\natexlab{c}}), \eprint{hep-ex/0403030}.

\bibitem[{\citenamefont{Aubert et~al.}(2005)}]{BABARSEMI}
\bibinfo{author}{\bibfnamefont{B.}~\bibnamefont{Aubert}} \bibnamefont{et~al.}
  (\bibinfo{collaboration}{\babar\ Collaboration}), \bibinfo{journal}{Phys.
  Rev.} \textbf{\bibinfo{volume}{D72}}, \bibinfo{pages}{052004}
  (\bibinfo{year}{2005}), \eprint{hep-ex/0508004}.

\bibitem[{\citenamefont{Aubert et~al.}()}]{BABARINCL}
\bibinfo{author}{\bibfnamefont{B.}~\bibnamefont{Aubert}} \bibnamefont{et~al.}
  (\bibinfo{collaboration}{\babar\ Collaboration}), \bibinfo{note}{(2005),
  hep-ex/0507001}.

\bibitem[{\citenamefont{Koppenburg et~al.}(2004)}]{BELLEBSG}
\bibinfo{author}{\bibfnamefont{P.}~\bibnamefont{Koppenburg}}
  \bibnamefont{et~al.} (\bibinfo{collaboration}{Belle Collaboration}),
  \bibinfo{journal}{Phys. Rev. Lett.} \textbf{\bibinfo{volume}{93}},
  \bibinfo{pages}{061803} (\bibinfo{year}{2004}), \eprint{hep-ex/0403004}.

\bibitem[{\citenamefont{Abe et~al.}()}]{BELLEBSGNEW}
\bibinfo{author}{\bibfnamefont{K.}~\bibnamefont{Abe}} \bibnamefont{et~al.}
  (\bibinfo{collaboration}{Belle Collaboration}), \bibinfo{note}{(2005),
  hep-ex/0508005}.

\bibitem[{\citenamefont{Acosta et~al.}(2005)}]{CDF}
\bibinfo{author}{\bibfnamefont{D.}~\bibnamefont{Acosta}} \bibnamefont{et~al.}
  (\bibinfo{collaboration}{CDF Collaboration}), \bibinfo{journal}{Phys. Rev.}
  \textbf{\bibinfo{volume}{D71}}, \bibinfo{pages}{051103}
  (\bibinfo{year}{2005}), \eprint{hep-ex/0502003}.

\bibitem[{\citenamefont{Csorna et~al.}(2004)}]{CLEOHAD}
\bibinfo{author}{\bibfnamefont{S.~E.} \bibnamefont{Csorna}}
  \bibnamefont{et~al.} (\bibinfo{collaboration}{CLEO Collaboration}),
  \bibinfo{journal}{Phys. Rev.} \textbf{\bibinfo{volume}{D70}},
  \bibinfo{pages}{032002} (\bibinfo{year}{2004}), \eprint{hep-ex/0403052}.

\bibitem[{\citenamefont{Chen et~al.}(2001)}]{CLEOBSG}
\bibinfo{author}{\bibfnamefont{S.}~\bibnamefont{Chen}} \bibnamefont{et~al.}
  (\bibinfo{collaboration}{CLEO Collaboration}), \bibinfo{journal}{Phys. Rev.
  Lett.} \textbf{\bibinfo{volume}{87}}, \bibinfo{pages}{251807}
  (\bibinfo{year}{2001}), \eprint{hep-ex/0108032}.

\bibitem[{\citenamefont{Alexander et~al.}(2005)}]{HFAG2004}
\bibinfo{author}{\bibfnamefont{J.}~\bibnamefont{Alexander}}
  \bibnamefont{et~al.} (\bibinfo{collaboration}{Heavy Flavor Averaging Group
  (HFAG)}) (\bibinfo{year}{2005}), \eprint{hep-ex/0412073}.

\bibitem[{\citenamefont{Abe et~al.}(2005{\natexlab{a}})}]{BELLEHAD}
\bibinfo{author}{\bibfnamefont{K.}~\bibnamefont{Abe}} \bibnamefont{et~al.}
  (\bibinfo{collaboration}{Belle Collaboration})
  (\bibinfo{year}{2005}{\natexlab{a}}), \eprint{hep-ex/0509013}.

\bibitem[{\citenamefont{Abe et~al.}(2005{\natexlab{b}})}]{BELLELEP}
\bibinfo{author}{\bibfnamefont{K.}~\bibnamefont{Abe}} \bibnamefont{et~al.}
  (\bibinfo{collaboration}{Belle Collaboration})
  (\bibinfo{year}{2005}{\natexlab{b}}), \eprint{hep-ex/0508056}.

\bibitem[{\citenamefont{Mahmood et~al.}(2004)}]{CLEOLEP}
\bibinfo{author}{\bibfnamefont{A.~H.} \bibnamefont{Mahmood}}
  \bibnamefont{et~al.} (\bibinfo{collaboration}{CLEO Collaboration}),
  \bibinfo{journal}{Phys. Rev.} \textbf{\bibinfo{volume}{D70}},
  \bibinfo{pages}{032003} (\bibinfo{year}{2004}), \eprint{hep-ex/0403053}.

\bibitem[{ste()}]{stepaniak}
\bibinfo{note}{A covariance matrix of the lepton energy moments can be found
  in: Ch. J. Stepaniak, Ph. D. thesis, University of Minnesota, (2004)}.

\bibitem[{\citenamefont{Trott}(2004)}]{Trott:2004xc}
\bibinfo{author}{\bibfnamefont{M.}~\bibnamefont{Trott}},
  \bibinfo{journal}{Phys. Rev.} \textbf{\bibinfo{volume}{D70}},
  \bibinfo{pages}{073003} (\bibinfo{year}{2004}), \eprint{hep-ph/0402120}.

\bibitem[{\citenamefont{Aquila et~al.}(2005)\citenamefont{Aquila, Gambino,
  Ridolfi, and Uraltsev}}]{Aquila:2005hq}
\bibinfo{author}{\bibfnamefont{V.}~\bibnamefont{Aquila}},
  \bibinfo{author}{\bibfnamefont{P.}~\bibnamefont{Gambino}},
  \bibinfo{author}{\bibfnamefont{G.}~\bibnamefont{Ridolfi}}, \bibnamefont{and}
  \bibinfo{author}{\bibfnamefont{N.}~\bibnamefont{Uraltsev}},
  \bibinfo{journal}{Nucl. Phys.} \textbf{\bibinfo{volume}{B719}},
  \bibinfo{pages}{77} (\bibinfo{year}{2005}), \eprint{hep-ph/0503083}.

\bibitem[{\citenamefont{Gambino}(2005)}]{Fortran}
\bibinfo{author}{\bibfnamefont{P.}~\bibnamefont{Gambino}}
  (\bibinfo{year}{2005}), \bibinfo{note}{private communication}.

\bibitem[{\citenamefont{Uraltsev}(2005{\natexlab{b}})}]{kolyapriv}
\bibinfo{author}{\bibfnamefont{N.}~\bibnamefont{Uraltsev}}
  (\bibinfo{year}{2005}{\natexlab{b}}), \bibinfo{note}{private communication}.

\bibitem[{\citenamefont{Hoang and Manohar}(2006)}]{Hoang:2005zw}
\bibinfo{author}{\bibfnamefont{A.~H.} \bibnamefont{Hoang}} \bibnamefont{and}
  \bibinfo{author}{\bibfnamefont{A.~V.} \bibnamefont{Manohar}},
  \bibinfo{journal}{Phys. Lett.} \textbf{\bibinfo{volume}{B633}},
  \bibinfo{pages}{526} (\bibinfo{year}{2006}), \eprint{hep-ph/0509195}.

\bibitem[{\citenamefont{Gray et~al.}(2005)}]{Gray:2005ur}
\bibinfo{author}{\bibfnamefont{A.}~\bibnamefont{Gray}} \bibnamefont{et~al.},
  \bibinfo{journal}{Phys. Rev.} \textbf{\bibinfo{volume}{D72}},
  \bibinfo{pages}{094507} (\bibinfo{year}{2005}), \eprint{hep-lat/0507013}.

\bibitem[{\citenamefont{Di~Renzo and Scorzato}(2004)}]{DiRenzo:2004xn}
\bibinfo{author}{\bibfnamefont{F.}~\bibnamefont{Di~Renzo}} \bibnamefont{and}
  \bibinfo{author}{\bibfnamefont{L.}~\bibnamefont{Scorzato}},
  \bibinfo{journal}{JHEP} \textbf{\bibinfo{volume}{11}}, \bibinfo{pages}{036}
  (\bibinfo{year}{2004}), \eprint{hep-lat/0408015}.

\bibitem[{\citenamefont{Nobes and Trottier}(2006)}]{Nobes:2005dz}
\bibinfo{author}{\bibfnamefont{M.}~\bibnamefont{Nobes}} \bibnamefont{and}
  \bibinfo{author}{\bibfnamefont{H.}~\bibnamefont{Trottier}},
  \bibinfo{journal}{PoS} \textbf{\bibinfo{volume}{LAT2005}},
  \bibinfo{pages}{209} (\bibinfo{year}{2006}), \eprint{hep-lat/0509128}.

\bibitem[{\citenamefont{Kagan and Neubert}(1999)}]{Kagan}
\bibinfo{author}{\bibfnamefont{A.~L.} \bibnamefont{Kagan}} \bibnamefont{and}
  \bibinfo{author}{\bibfnamefont{M.}~\bibnamefont{Neubert}},
  \bibinfo{journal}{Eur. Phys. J.} \textbf{\bibinfo{volume}{C7}},
  \bibinfo{pages}{5} (\bibinfo{year}{1999}), \eprint{hep-ph/9805303}.

\bibitem[{\citenamefont{Bosch et~al.}(2004{\natexlab{a}})\citenamefont{Bosch,
  Lange, Neubert, and Paz}}]{Bosch:2004th}
\bibinfo{author}{\bibfnamefont{S.~W.} \bibnamefont{Bosch}},
  \bibinfo{author}{\bibfnamefont{B.~O.} \bibnamefont{Lange}},
  \bibinfo{author}{\bibfnamefont{M.}~\bibnamefont{Neubert}}, \bibnamefont{and}
  \bibinfo{author}{\bibfnamefont{G.}~\bibnamefont{Paz}},
  \bibinfo{journal}{Nucl. Phys.} \textbf{\bibinfo{volume}{B699}},
  \bibinfo{pages}{335} (\bibinfo{year}{2004}{\natexlab{a}}),
  \eprint{hep-ph/0402094}.

\bibitem[{\citenamefont{Lange}(2001)}]{Lange:2001uf}
\bibinfo{author}{\bibfnamefont{D.~J.} \bibnamefont{Lange}},
  \bibinfo{journal}{Nucl. Instrum. Meth.} \textbf{\bibinfo{volume}{A462}},
  \bibinfo{pages}{152} (\bibinfo{year}{2001}).

\bibitem[{\citenamefont{Neubert}(2005{\natexlab{a}})}]{Neubert:2004sp}
\bibinfo{author}{\bibfnamefont{M.}~\bibnamefont{Neubert}},
  \bibinfo{journal}{Phys. Lett.} \textbf{\bibinfo{volume}{B612}},
  \bibinfo{pages}{13} (\bibinfo{year}{2005}{\natexlab{a}}),
  \eprint{hep-ph/0412241}.

\bibitem[{\citenamefont{Lange et~al.}(2005)\citenamefont{Lange, Neubert, and
  Paz}}]{Lange:2005yw}
\bibinfo{author}{\bibfnamefont{B.~O.} \bibnamefont{Lange}},
  \bibinfo{author}{\bibfnamefont{M.}~\bibnamefont{Neubert}}, \bibnamefont{and}
  \bibinfo{author}{\bibfnamefont{G.}~\bibnamefont{Paz}},
  \bibinfo{journal}{Phys. Rev.} \textbf{\bibinfo{volume}{D72}},
  \bibinfo{pages}{073006} (\bibinfo{year}{2005}), \eprint{hep-ph/0504071}.

\bibitem[{\citenamefont{Neubert}(2005{\natexlab{b}})}]{Neubert:2004cu}
\bibinfo{author}{\bibfnamefont{M.}~\bibnamefont{Neubert}},
  \bibinfo{journal}{Eur. Phys. J.} \textbf{\bibinfo{volume}{C44}},
  \bibinfo{pages}{205} (\bibinfo{year}{2005}{\natexlab{b}}),
  \eprint{hep-ph/0411027}.

\bibitem[{\citenamefont{Bosch et~al.}(2004{\natexlab{b}})\citenamefont{Bosch,
  Neubert, and Paz}}]{Bosch:2004cb}
\bibinfo{author}{\bibfnamefont{S.~W.} \bibnamefont{Bosch}},
  \bibinfo{author}{\bibfnamefont{M.}~\bibnamefont{Neubert}}, \bibnamefont{and}
  \bibinfo{author}{\bibfnamefont{G.}~\bibnamefont{Paz}},
  \bibinfo{journal}{JHEP} \textbf{\bibinfo{volume}{11}}, \bibinfo{pages}{073}
  (\bibinfo{year}{2004}{\natexlab{b}}), \eprint{hep-ph/0409115}.

\bibitem[{\citenamefont{Neubert}(2005{\natexlab{c}})}]{Neubert:2004dd}
\bibinfo{author}{\bibfnamefont{M.}~\bibnamefont{Neubert}},
  \bibinfo{journal}{Eur. Phys. J.} \textbf{\bibinfo{volume}{C40}},
  \bibinfo{pages}{165} (\bibinfo{year}{2005}{\natexlab{c}}),
  \eprint{hep-ph/0408179}.

\bibitem[{\citenamefont{Uraltsev}(1999)}]{Uraltsev:1999rr}
\bibinfo{author}{\bibfnamefont{N.}~\bibnamefont{Uraltsev}},
  \bibinfo{journal}{Int. J. Mod. Phys.} \textbf{\bibinfo{volume}{A14}},
  \bibinfo{pages}{4641} (\bibinfo{year}{1999}), \eprint{hep-ph/9905520}.

\bibitem[{\citenamefont{Gambino and Misiak}(2001)}]{Gambino:2001ew}
\bibinfo{author}{\bibfnamefont{P.}~\bibnamefont{Gambino}} \bibnamefont{and}
  \bibinfo{author}{\bibfnamefont{M.}~\bibnamefont{Misiak}},
  \bibinfo{journal}{Nucl. Phys.} \textbf{\bibinfo{volume}{B611}},
  \bibinfo{pages}{338} (\bibinfo{year}{2001}), \eprint{hep-ph/0104034}.

\bibitem[{\citenamefont{Buras et~al.}(2002)\citenamefont{Buras, Czarnecki,
  Misiak, and Urban}}]{Buras:2002tp}
\bibinfo{author}{\bibfnamefont{A.~J.} \bibnamefont{Buras}},
  \bibinfo{author}{\bibfnamefont{A.}~\bibnamefont{Czarnecki}},
  \bibinfo{author}{\bibfnamefont{M.}~\bibnamefont{Misiak}}, \bibnamefont{and}
  \bibinfo{author}{\bibfnamefont{J.}~\bibnamefont{Urban}},
  \bibinfo{journal}{Nucl. Phys.} \textbf{\bibinfo{volume}{B631}},
  \bibinfo{pages}{219} (\bibinfo{year}{2002}), \eprint{hep-ph/0203135}.

\end{thebibliography}

\end{document}